%% file: Stability_Map__Baaser__2025__JoE_Revision_A.tex
\documentclass{article}
\usepackage{graphicx}
\usepackage{color}           
\newcommand{\dint}{{\rm d}}
\newcommand{\Dabl}{{\rm D}}                    
\usepackage{epsfig}
\usepackage{amsmath}          
\usepackage{amssymb}
\usepackage{latexsym}
\def\pb#1{\boldsymbol{#1}}            

\def\f #1{\hbox{\bf #1}}

\begin{document}

\title{Hyperelastic stability landscape:
A check for {\sc Hill} stability of isotropic, incompressible hyperelasticity
depending on material parameters}


\author{Herbert Baaser, University of Applied Sciences Bingen, Germany}


%
%
\maketitle
\begin{abstract}
In this paper, we describe a uniform and standardized approach for analytically verifying the stability of isotropic, incompressible hyperelastic material models.
Here, we address {\sl stability} as fulfillment of the {\sc Hill} condition -- i.e.\ the positive definiteness of the material modulus in the {\sc Kirchhoff} stress -- log--strain relation.
For incompressible material behavior, all mathematically and mechanically possible deformations lie within a range bounded, on the one hand, by uniaxial states and, on the other hand, by biaxial states; shear {deformation} states lie in between. This becomes particularly clear when the possible states are represented in the invariant plane. This very representation is now also used to visualize the regions of unstable material behavior depending on the selected strain energy function and the respective data set of material parameters.
This demonstrates how, for some constellations of energy functions, with appropriate selection or calibration of parameters, stable and unstable regions can be observed. If such cases occur, it is no longer legitimate to use them to initiate, for example, finite element simulations. This is particularly striking when, for example, a fit appears stable in uniaxial tension, but the same parameter set for shear states results in unstable behavior without this being specifically investigated.
The presented approach can reveal simple indicators for this.
{Nevertheless, the {\sl simple shear} deformation, where the principal axes lag behind the deformation $\gamma=\tan\alpha$ of the shear angle $\alpha$, i.e.\ the rotation tensor $\f R \neq \f I$, still represents a special case that requires extra investigations. This is especially true given that all shear components of the logarithmic strains themselves exhibit a non--monotonic behavior with respect to the deformation angle.}
%
%
\end{abstract}
\section{Introduction \label{intro}}
The representation of classical isotropic hyperelasticity as a function of the input variables in the form of deformation invariants, see \cite{Hol00},
or of the stretches themselves, see \cite{Ogd72}, is still of importance, especially when -- as here -- stability statements of the
models are concerned when used in the context of the finite element method (FEM).
The special case of {\sl quasi--}incompressibility is treated as a constraint within the framework of FEM, see modern textbooks such as \cite{Wri08} or more recent literature, but it is very well suited in this context to analytically describe the procedure and to identify specific properties of the models.
{In this context of material parameter calibration, we limit ourselves exclusively to ideal--incompressible deformation states.}
Elastomers in technical products for vibration control and sealing technology are frequently mentioned as a class of materials that exhibit almost ideal incompressibility.
%
%

This work can be read in series with \cite{CrHuDoHu00} and \cite{DiGi05}, where "three natural strain invariants"
{$K_1=\log J$ as "amount--of--dilatation", $K_2=\Vert{\rm dev} \log \f v\rVert$ as "magnitude--of--distortion" and $K_3=\frac{3\sqrt{6}}{K_2}\det({\rm dev}(\log \f v))$ as "mode--of--distortion"}
are proposed based on the left polar decomposition of the deformation gradient $\f F=\f v\cdot\f R$, see \cite{NeLaMa24}, with $J=\det\f F$
in order to specify hyperelastic deformation modes alternatively.
In that sense, one can distinguish the {\sl prototype deformation modes} from each other by $K_3=0$ for pure shear, $K_3=1$ for uniaxial tension and $K_3=-1$ for equibiaxial tension.

In a similar way, it was also possible in \cite{BaHoSc12} to find a formulation that depends significantly only on the stretch in the tensile direction and is used here because this representation
allows a further derivation of the stress in a simple way. {Starting from the general representation of the standard invariants
\begin{eqnarray}
I_1 &=& \lambda_1^2+\lambda_2^2+\lambda_3^2 \,, \\
I_2 &=& \lambda_1^2\lambda_2^2+\lambda_1^2\lambda_3^2+\lambda_2^2\lambda_3^2 \,, \nonumber \\
I_3 &=& \lambda_1^2\, \lambda_2^2\, \lambda_3^2 =: J^2\nonumber
\end{eqnarray}
as a function of the stretches $\lambda_{1,2,3}$ and its specification for the incompressible case with $J:=\det \f F \equiv 1$ and {$\lambda_3=(\lambda_1\, \lambda_2)^{-1}$} as
\begin{equation}
\bar{I}_1=\lambda_1^2+\lambda_2^2+\frac{1}{\lambda_1^2\, \lambda_2^2} \quad {\rm and} \quad \bar{I}_2=\lambda_1^2\lambda_2^2+\frac{1}{\lambda_1^2}+\frac{1}{\lambda_2^2}
\label{Invar_1_2}
\end{equation}
we directly obtain}
\begin{equation}
\bar{I}_{1,m}=\lambda^2+\lambda^{2m}+\lambda^{-2m-2} \quad {\rm and} \quad \bar{I}_{2,m}=\lambda^{-2}+\lambda^{-2m}+\lambda^{2m+2}
\label{Invar_lam_m}
\end{equation}
as functions of an (uniaxial) stretch (intensity) $\lambda$ and an additional parameter $m=[-\frac12 ... 0 ... 1]$ to specify the
deformation mode {\sl uniaxial tension} {($m=-\frac12$)}, {\sl pure shear} {($m=0$, often called {\sl plane strain})} and {\sl equibiaxial tension} {($m=1$)}, respectively.
%
%
This ends up in the representation of the modes as boundary lines of all possible deformations in case of
incompressibility in the {\sl plane of invariants}, following e.g.\ \cite{Tre75}. In the Appendix \ref{app-B_boundarylines} we give more details on this graphical representation.

{In \cite{AnGoSa24} and recently in \cite{HoMu25} based on \cite{YaChHu22},
a very useful representation of the three squared stretches $\lambda_1^2, \lambda_2^2, \lambda_3^2$ as functions of the classical invariants
$I_1, I_2$ is given for incompressible hyperelasticity in analytical form, i.e.\ $J=\lambda_1 \lambda_2 \lambda_3 \equiv 1$ or $I_3=J^2=1$, equivalently.
In addition, here we give an equivalent representation and conversion of any given incompressible deformation state $\lambda_1^2, \lambda_2^2, \lambda_3^2$ or
$\bar{I}_1, \bar{I}_2$ into the ($\lambda, m$) representation given above,
see Appendix \ref{app-A} for a short description and derivation of that closed form representation --- to the knowledge of the author for the first time.}
In consequence, any given representation $\lambda_1^2, \lambda_2^2, \lambda_3^2$ {with $\lambda_1\lambda_2\lambda_3\equiv 1$} or $\bar{I}_1, \bar{I}_2$ can
be converted into each other, in both directions, as needed.
In this series there are also new investigations \cite{Kos25} on the stability specifically for the {\sc Mooney--Rivlin} model, which is also used here,
where, among other things, the relationship between the two material parameters of this model is investigated.
There, stability in the three prototype deformation modes from above is defined as positive slope of the {\sc Cauchy} stress versus the relevant stretch $\lambda$,
i.e.\ the {\sc Cauchy} stress--stretch curve must be monotone increasing.

Here, an analogue representation of isotropic hyperelasticity is given in a general form,
in the sense that the three prototype deformations (uniaxial tension, shear and equibiaxial tension) are formulated uniformly based on (\ref{Invar_lam_m}).
From this, a map of possible regions of material stability is developed,
following the stabi\-lity criteria in modern FEM codes such as {\sc Abaqus} or {\sc Ansys}, see \cite{Aba20}: the application of the {\sc Hill} stability
criterion\footnote{In {\sc Abaqus} denoted vaguely as material stability or {\sc Drucker} stability criterion with respect to the references given therein.
The {\sc Drucker} criterion is best known in {\sl geometrically linearized elasticity} as $\dint\pb\sigma:\dint\pb\varepsilon>0 \Longleftrightarrow
\left<\pb\sigma( \pb\varepsilon_1)-\pb\sigma( \pb\varepsilon_2), \pb\varepsilon_1-\pb\varepsilon_2  \right> >0
\quad \forall\, \pb\varepsilon_1\neq \pb\varepsilon_2$ expressing the convexity of the strain energy in the infinitesimal strain
$\varepsilon=\frac12[\nabla\f u+(\nabla\f u)^{\rm T}]$ as symmetrized gradient of the field of displacements $\f u=[u_1\, u_2\, u_3]^{\rm T}$
with $\nabla\f u=\frac{\partial u_i}{\partial x_j}$ in three--dimensional space ($i,j=1,2,3$).
For linear isotropy, this {\sc Drucker} criterion amounts to $\mu>0, \kappa>0$, where $\mu$ is the shear modulus, and $\kappa$ is the bulk modulus.}
for general, incompressible hyperelastic models formulated as a potential function in $\bar{I}_1$ and $\bar{I}_2$ or equivalently in $\lambda_{1,2,3}$, as already given in \cite{Dru59} based on \cite{Hil58}.
It must be mentioned that deformations with rotating principal axes, such as {\sl simple shear}, are indeed included in the consideration of the plane of invariants (see \cite{BaHoSc12}),
but cases of non--monotone {\sc Cauchy} stress behavior can certainly be shown there, which will be discussed further, see \cite{NeHuNgKoMa25}, \cite{WoHoNe25} and a simple example in Appendix \ref{app-B_SiSh}.

The aim of this work is to consistently apply these findings to verify the stability of the material formulation
and to visualize it in a suitable mathematical--mechanical space.
This leads to the graphical representation of the results in the {\sl plane of invariants}, see \cite{Tre75}, showing systematically the stability of material models
based on their parameters.
We consider a strictly one--dimensional stability marker $D_{1D}^m>0$, generalizing the {\sc Cauchy} stress monotonicity from the three prototype modes to
the whole plane of invariants and the full {\sc Hill's} inequality for the incompressible case.
To this end, we derive a concise formulation of {\sc Hill's} inequality based on the positive definiteness of a projected {\sc Hesse} matrix.
These methods are both analytically and numerically verifiable and directly correspond to typical requirements in finite elements codes such as {\sc Abaqus} or {\sc Ansys}.
The application of the formulation used here with tools for the automatic,
analytical derivation of arbitrary energy potentials allows this graphical representation in, for example, {\sc Matlab} or {\sc Excel} as a function
of the material parameters.
\section{Background of stability analysis \label{sec:Background}}
We denote the first {\sc Piola--Kirchhoff} stress tensor by $\f S_1$ and $\f F=\partial \f x/\partial \f X$ the deformation gradient between the actual
configuration $\f x$ and the reference configuration $\f X$. Thus, the mechanical power ${\mathcal P} = {\f S}_1 : \dot{\f F}$ results from the rate of the deformation gradient $\f F$.
Please note in this context the identity of the true {\sc Cauchy} stress $\pb \sigma$ and the
{\sc Kirchhoff} stress $\pb \tau = J \pb \sigma = \f S_1 \cdot \f F^{\rm T}$ in the case of incompressibility, i.e.\ $J=\det \f F \equiv 1$.
Our investigations follow recent works such as \cite{AgHoBeSkGhMaNe25} and \cite{GrScMaNe2019} developing \cite{Hil68},\cite{Sid74} and \cite{Hil70},
where {\sl the notion of conjugate pairs of stress and strain measures} [...] is introduced.
Already {\sc Richter} \cite{Ric48} uses a logarithmic strain matrix is used, in which the separation into volumetric and isochoric parts is realized by a deviatoric representation.
Thus, following \cite{Ric49} the simplest definition of the strain tensor appears to be the logarithmic deformation matrix.

The fundamental recent investigation in \cite{NeHuNgKoMa25} gives a theoretical underpinning of the natural appearance of the logarithmic strains based on
the "corotational stability postulate" (CSP), i.e.\ for any reasonable corotational rate $\frac{\Dabl^\circ}{\Dabl t}$ of the {\sc Cauchy} stress $\pb\sigma$ it applies
\begin{equation}
\left<\dfrac{\Dabl^\circ}{\Dabl t}\pb\sigma, \f d\right> >0 \quad \forall\, \f d\neq\f 0
\label{postulate_CSP}
\end{equation}
with $\f d:=\frac12(\f l+\f l^{\rm T})$ as symmetric part of the spatial velocity gradient $\f l:={\rm grad}\,\dot{\f x}=\dot{\f F}\cdot\f F^{-1}$.
The {\sc Zaremba-Jaumann} rate $\frac{\Dabl^{ZJ}}{\Dabl t}$ is one prototype corotational rate.
It is further motivated, that {\sl the corotational stability postulate (CSP) is a reasonable constitutive stability postulate in nonlinear elasticity, complementing local material stability viz.\
{\sc Legendre--Hadamard}--ellipticity.}
In the case of compressibility, not treated here, condition (\ref{postulate_CSP}) is fully equivalent to
\begin{eqnarray}
\label{stabil_compress}
\dint \pb \sigma : \dint \pb \epsilon^{\rm log} &=& \left<\dint\pb\sigma, \dint\pb\epsilon^{\rm log}\right> \stackrel{!}{>} 0 \\
                                                                           &=&  \dint\sigma_1\,\dint\epsilon_1^{\rm log} + \dint\sigma_2\,\dint\epsilon_2^{\rm log} + \dint\sigma_3\,\dint\epsilon_3^{\rm log} >0 \,\, \forall\, \f v_1 \neq \f v_2 \nonumber \Longleftrightarrow
\end{eqnarray}
\begin{equation*}
\left<\pb\sigma(\log \f v_2) -\pb\sigma(\log \f v_1),\, \log(\f v_2)-\log(\f v_1)\right> >0\,\, \forall\, \f v_1 \neq \f v_2
\end{equation*}
and amounts to the fundamental {\sl True--Stress--True--Strain Monotonicity} condition ("TSTS--M$^{+}$") proposed for study
of elastic stability in \cite{NeHoAgBeSkGhMa25} and \cite{NeHuNgKoMa25}. 

Given that and following \cite{Hil58,Hil68,Dru59,Sid74}, "{\sc Hill} stability" for the incompressible case can be expressed as
\begin{equation}
\dint \pb \sigma : \dint \pb \epsilon^{\rm log} = \left<\dint\pb\sigma, \dint\pb\epsilon^{\rm log}\right>\stackrel{!}{>} 0\,, \quad {\rm trace}(\dint\pb\epsilon^{\rm log})=0  \quad\Longleftrightarrow
\label{Drucker}
\end{equation}
\begin{equation*}
\left<\pb\sigma(\log \f v_2) -\pb\sigma(\log \f v_1),\, \log(\f v_2)-\log(\f v_1)\right> >0\,\, \forall\, \f v_1 \neq \f v_2 \, , \det\f v_1=\det\f v_2=1\,,
\end{equation*}
where $\dint \pb \sigma$ denotes the increment of true {\sc Cauchy} stress\footnote{In former {\sc Abaqus} implementations (see \cite{Aba09})
the increments of the {\sc Kirchhoff} stress $\dint \pb \tau$ are used, whereas now (see \cite{Aba20}) the relation (\ref{D_mat}) seems to be used.
In \cite{NeHoAgBeSkGhMa25} it has been shown that (\ref{Drucker}) coincides with the {\sl corotational stability postulate (CSP)} for incompressibility.
Nonetheless, to the understanding of the author, {\sc Abaqus} and {\sc Ansys} use {\sc Hill's} condition also in the compressible case as stability indicator.
Then, {\sc Hill's} condition is not equivalent to TSTS--M$^{+}$ (\ref{stabil_compress}) and looses it's physical significance.\label{footnote_ABAQ_Drucker}},
$\pb \epsilon^{\rm log}$ the logarithmic {\sc Hencky} strain and "$:$" the double contracting tensor
product, i.e.\ a scalar re\-pre\-sentation $s=A_{ij}B_{ij}$ for second order tensors $\f A$ and $\f B$;
$\f v_1$ and $\f v_2$ with $\f v=\f F\cdot \f R^{-1}$ denote any left stretch tensors, see also \cite{WoHoNe25} recently.

Since $\pb\tau=\pb\sigma$ for incompressibility, that means, as in \cite{Hil68}, {\sc Hill} stability is more precisely associated with the statement
\begin{equation}
\dint \pb\tau : \dint\pb\epsilon^{\rm log} > 0\,, \quad {\rm trace}(\dint\pb\epsilon^{\rm log})=0 \quad\Longleftrightarrow
\label{Drucker_Hill}
\end{equation}
\begin{equation*}
\left<\pb\tau(\log \f v_2) -\pb\tau(\log \f v_1),\, \log(\f v_2)-\log(\f v_1)\right> >0\,\, \forall \f v_1 \neq \f v_2 \, , \det\f v_1=\det\f v_2=1\,,
\end{equation*}
which implies that the deviatoric part of the {\sc Kirchhoff} stress $\pb\tau$ is a monotone function of $\log \f v$. The latter statement is the generic condition of {\sc Hill} stability for compressible solids.

In modern FE--codes, the evaluation of (\ref{Drucker_Hill}) seems to be the crucial point to check or guarantee for material stability; 
\begin{figure}[!hbpt]
\centering
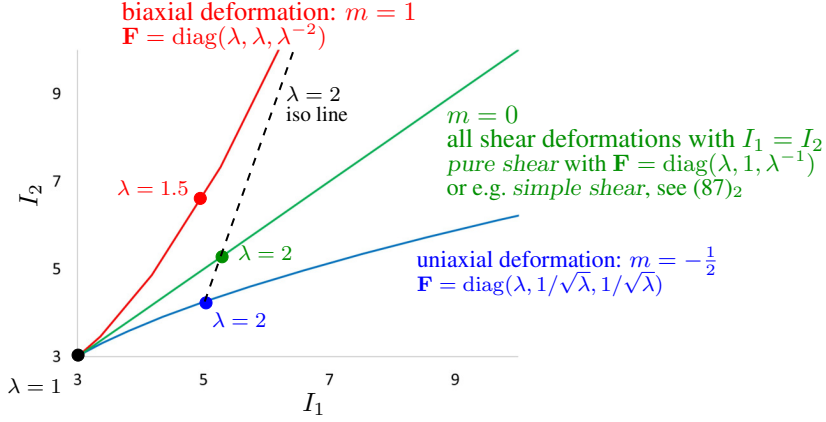  
\caption{Illustration of {\sc Abaqus} check for stability only for the three {special} deformation modes in the plane of invariants up to $\lambda_\star=10$.
In addition, we give the $\lambda=2$--iso line in that diagram exemplary for all $\lambda=const$--iso lines as straight lines across that plane.
Note that a proof is still pending that monotonicity of the {\sc Kirchhoff} stress $\pb\tau$ with respect to $\log\f v$ in the three prototype deformation modes simultaneously implies monotonicity in all deformations without rotating principal
axes, i.e.\ where $\f F$ is diagonal.
However, {\sc Hill}'s condition (\ref{Drucker_Hill}) implies for all deformation modes without rotating principal axes, that the relevant stress curve must be monotone, see \cite{NeHuNgKoMa25}.
Here, we will only check the above monotonicity as a necessary condition for (\ref{Drucker}).}
\label{Abaq_Check_InvarEbene}
\end{figure}
in the sense of (\ref{Drucker})$_1$ as given in {\sc Abaqus} {\sl User's Manual} \cite{Aba20}, Sec.\ 21.5.1, the check of {\sl stability is enforced for the three [here called "prototype"] deformation modes uniaxial tension, equi\-bi\-axial tension, planar tension}, i.e.\ pure shear, performing it {\sl for the polynomial models, the {\sc Ogden} potential, the {\sc Van der Waals} form [...].}
This check is executed only for each model once before the simulation is started individually in tension and compression direction,
for $\lambda = 0.1$ up to $\lambda = 10$ by increments of $\Delta \lambda = 0.01$.
In the {\sc Ansys} {\sl Mechanical APDL Theory Manual} \cite{Ans25} [Section 4.6.9., eqn.\ (4.225)] the same statements are given concerning the {\sc Mooney--Rivlin}, the {\sc Ogden}
and the polynomial form hyperelastic material models, i.e.\  the check is performed for the {\sl Kirchhoff stress tensor corresponding to a change in the logarithmic
strain}.

In the Appendix~\ref{app-B_boundarylines} we give the analytical derivation of the function {$\bar{I}_2^{\rm uniax}=\bar{I}_2^{\rm uniax}(\bar{I}_1^{\rm uniax})$} exemplary for the
here given uniaxial deformation mode in \textcolor{blue}{blue}, see Fig.\ \ref{Abaq_Check_InvarEbene}.
The corresponding boundary line for the biaxial deformation mode (here in \textcolor{red}{red}) is obtained in an analogous way and represents the reflection of the curve at the angle bisector, the \textcolor{green}{green} line of the shear deformation modes.

\begin{figure}[!hbpt]
\centering
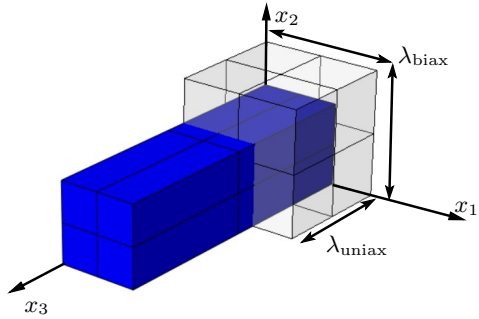
\caption{A biaxial deformation along $x_1$ and $x_2$ with $\lambda_{\rm biax}>1$ can be interpreted (and experimentally observed) as uniaxial compression in $x_3$ direction with $\lambda_3=\lambda_{\rm uniax}<1$.}
\label{biax_uniax_equivalence}
\end{figure}

\begin{figure}[!hbpt]
\centering
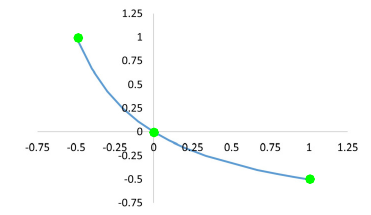
\caption{Equivalence of $m$ and $m^\star$ in the sense of a change of sign in tension ($\lambda>1$) and compression ($\lambda<1$).
Examplary, $m=1$ for the biaxial deformation mode corresponds to $m^\star=-0.5$ as uniaxial mode, see Appendix \ref{app-B_boundarylines} and Fig.\ \ref{invar_m4}.}
\label{m_equiv}
\end{figure}

In addition, we remark that a biaxial deformation along $x_1$ and $x_2$ with $\lambda_{\rm biax}>1$ can be interpreted (and experimentally observed) as uniaxial compression with $\lambda_{\rm uniax}<1$ and vice versa, see Fig.\ \ref{biax_uniax_equivalence} for illustration.
%
%
Furthermore, in Fig. \ref{m_equiv} we specify the relationship between "equivalent" mode parameters $m$ and $m^\star$.
This equivalence describes the sign change from positive to negative $m$ and vice versa.
%
This can be seen as proof that the \textcolor{blue}{blue} and \textcolor{red}{red} lines in Fig.\ \ref{Abaq_Check_InvarEbene} are each reflected across the (\textcolor{green}{green}) angle bisector.
However, this does not mean that iso-lines for "m" are also reflected across the angle bisector.
In the Appendix \ref{app-A} and the Appendix \ref{app-B_boundarylines} we give more details and further insights into possible derivations for these connections
of $m$ and $m^\star$ for $\lambda>1$ and $\lambda<1$.

As denoted furtheron in {\sc Abaqus} {\sl User's Manual} \cite{Aba20}, Sec.\ 21.5.1, {\sl "the material is assumed to be incompressible,
[and we] can choose any value for the hydrostatic pressure without affecting the strains.  A convenient choice for the stability calculation is $\sigma_3=\dint \sigma_3=0$.}"
So, (\ref{Drucker}) is reformulated in the case of incompressible isotropy and in terms of the principal stresses and strains to 
\begin{equation}
\begin{bmatrix}\dint \tau_1 \\ \dint \tau_2\end{bmatrix}
=\begin{bmatrix}\dint \sigma_1 \\ \dint \sigma_2\end{bmatrix}
=\underbrace{\begin{bmatrix}D_{11} \quad D_{12}\\ D_{21} \quad D_{22}\end{bmatrix}}_{=:\mathbb D}\cdot\begin{bmatrix}\dint \epsilon^{\rm log}_1 \\ \dint \epsilon^{\rm log}_2\end{bmatrix}
\label{D_mat}
\end{equation}
with symmetric material stiffness tensor
\begin{eqnarray}
\mathbb D&=&\begin{bmatrix}\Dabl_{\log \lambda_1}\, \tau_1(\log \lambda_1,\log \lambda_2) \quad \Dabl_{\log \lambda_2}\, \tau_1(\log \lambda_1,\log \lambda_2) \\ 
                                                   \Dabl_{\log \lambda_1}\, \tau_2(\log \lambda_1,\log \lambda_2) \quad \Dabl_{\log \lambda_2}\, \tau_2(\log \lambda_1,\log \lambda_2) \end{bmatrix} \in{\rm Sym}(2)\nonumber\\
&=&\begin{bmatrix}\lambda_1\Dabl_{\lambda_1}\, \tau_1(\log \lambda_1,\log \lambda_2) \quad \lambda_2\Dabl_{\lambda_2}\, \tau_1(\log \lambda_1,\log \lambda_2) \\ 
                                 \lambda_1\Dabl_{\lambda_1}\, \tau_2(\log \lambda_1,\log \lambda_2) \quad \lambda_2\Dabl_{\lambda_2}\, \tau_2(\log \lambda_1,\log \lambda_2) \end{bmatrix}
\end{eqnarray}
%
%
%
respecting for $\dfrac{\dint \lambda_i}{\dint\log\lambda_i}=\lambda_i$ for $i=1,2$, see (\ref{log_eps_inv}) lateron. Alternative representations of $\mathbb D$ are presented in the
Appendix in (\ref{app_D_altern}) and (\ref{def_W__inc_red}).
The stability is then checked by the positive definiteness of the tangent stiffness matrix $\mathbb D$ with
\begin{equation}
{\rm trace}(\mathbb D) = D_{11}+D_{22}>0 \quad{\rm and}\quad \det\,\mathbb D=D_{11}\, D_{22}-D_{12}\,D_{21}>0\,,
\end{equation}
see \cite{Aba20}, whereas the classical conditions $D_{11}>0$ and $\det\, \mathbb D>0$ for positive defi\-nite\-ness of $\mathbb D$ are indeed equivalent for symmetric $2\times 2$ matrices.
Summarizing, the built--in {\sc Abaqus} stability routine checks the positive definiteness of $\mathbb D$ in (\ref{D_mat}) only along the uniaxial tension,
the equibiaxial tension and the pure shear paths,
see Fig.~\ref{Abaq_Check_InvarEbene} and Fig.\ \ref{InvarEbene_Yeoh_Stab__ABAQ} in Section~\ref{sec:ABAQ_check} below.
Since the check of positive definiteness is given for a symmetric tensor $\mathbb D$, this shows that {\sc Abaqus} is really only considering {\sc Hill’s} inequality
as mentioned in the footnote \ref{footnote_ABAQ_Drucker} before.
For that, in the Appendix \ref{app-C_ABAQ_stabi} we give the {\sc Matlab} code for the evaluation of (\ref{D_mat}) exemplary for the {\sc Yeoh} example resulting in Fig.\ \ref{InvarEbene_Yeoh_Stab__ABAQ}.
%
%
%
%
%

The evaluation of the principal stress components follows directly the representation of \cite{Hol00}, p.\ 225 (6.69), given as
\begin{equation}
\sigma_a  = -p+\lambda_a\left(\frac{\dint W}{\dint \bar{I}_1}\,\frac{\dint \bar{I}_1}{\dint \lambda_a}+\frac{\dint W}{\dint \bar{I}_2}\,\frac{\dint \bar{I}_2}{\lambda_2}\right)                                
\label{sig12_HA}
\end{equation}
for $a=1,2,3$, whereas $\sigma_3=\tau_3=0$ ("plane stress") from which we obtain
\begin{equation}
p=\lambda_3\left(\frac{\dint W}{\dint \bar{I}_1}\,\frac{\dint \bar{I}_1}{\dint \lambda_3}+\frac{\dint W}{\dint \bar{I}_2}\,\frac{\dint \bar{I}_2}{\lambda_3}\right)
\end{equation}
for the still open hydrostatic pressure component in (\ref{sig12_HA}).
Reordering (\ref{sig12_HA}) and inserting $p$ leads to
\begin{eqnarray}
\tau_1 &=& \sigma_1 = 2(\lambda^2-\lambda^{-2m-2})\frac{\dint W}{\dint \bar{I}_1} + 2(\lambda^{2m+2}-\lambda^{-2})\frac{\dint W}{\dint \bar{I}_2}\,, \nonumber\\
\tau_2 &=& \sigma_2 = 2(\lambda^{2m}-\lambda^{-2m-2})\frac{\dint W}{\dint \bar{I}_1} + 2(\lambda^{2m+2}-\lambda^{-2m})\frac{\dint W}{\dint \bar{I}_2}
\label{tau_1__tau_2}
\end{eqnarray}
in general for the two principal components at plane stress and incompressibility conditions, which now can be specified along the three paths of Fig.\ \ref{Abaq_Check_InvarEbene}
leading each to
\begin{itemize}
\item[$\diamond$] $\lambda_1=\lambda, \lambda_2=\lambda_3=\lambda^{-\frac12}$ for {\sl uniaxial tension} with $m=-\frac12$,
\begin{eqnarray}
\sigma_1 &=& 2(\lambda^2-\lambda^{-1})\frac{\dint W}{\dint \bar{I}_1} + 2(\lambda-\lambda^{-2})\frac{\dint W}{\dint \bar{I}_2}\,, \nonumber\\
\sigma_2 &=& 0
\label{tau_1__tau_2__1ax}
\end{eqnarray}
\item[$\diamond$] $\lambda_1=\lambda, \lambda_2=1, \lambda_3=\lambda^{-1}$ for {\sl pure shear} or {\sl plane strain mode} with $m=0$
\begin{eqnarray}
\sigma_1 &=& 2(\lambda^2-\lambda^{-2})(\frac{\dint W}{\dint \bar{I}_1}+\frac{\dint W}{\dint \bar{I}_2})\,, \nonumber\\
\sigma_2 &=& 2(1-\lambda^{-2})\frac{\dint W}{\dint \bar{I}_1} + 2(\lambda^2-1)\frac{\dint W}{\dint \bar{I}_2}
\label{tau_1__tau_2__PS}
\end{eqnarray}
\item[$\diamond$] $\lambda_1=\lambda_2=\lambda, \lambda_3=\lambda^{-2}$ for {\sl biaxial tension} with $m=1$, respectively,
\begin{eqnarray}
\sigma_1 &=& 2(\lambda^2-\lambda^{-4})\frac{\dint W}{\dint \bar{I}_1} + 2(\lambda^4-\lambda^{-2})\frac{\dint W}{\dint \bar{I}_2}\,, \nonumber\\
\sigma_2 &=& 2(\lambda^2-\lambda^{-4})\frac{\dint W}{\dint \bar{I}_1} + 2(\lambda^4-\lambda^{-2})\frac{\dint W}{\dint \bar{I}_2} = \sigma_1
\label{tau_1__tau_2__biax}
\end{eqnarray}
\end{itemize}
in equivalence to \cite{Aba20}.

\begin{figure}[!hb]
\centering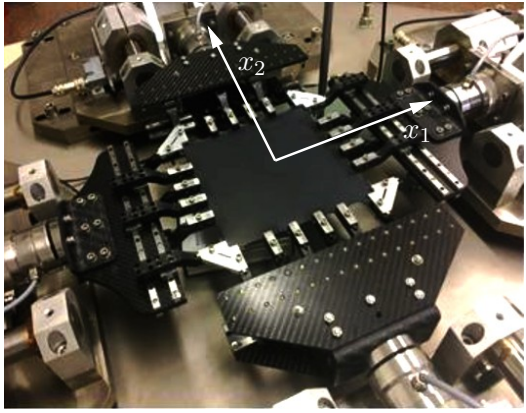
\caption{Possible experimental setup by a "biax--tester" from \cite{ScCaLoKiHoScDeHe17}, where in the thickness direction $x_3$ an uniaxial compression now takes place, see Fig.\ \ref{biax_uniax_equivalence}.}
\label{biax-tester}
\end{figure}

In addition, we give the resulting principal stresses (\ref{tau_1__tau_2}) exemplary for an arbitrary value of the mode parameter $m$, here e.g.\ $m=\frac13$ as depicted lateron in Fig.\ \ref{InvarEbene_Yeoh_Stab}:
\begin{eqnarray}
\sigma_1 &=& 2(\lambda^2-\lambda^{-\frac83})\frac{\dint W}{\dint \bar{I}_1} + 2(\lambda^\frac83-\lambda^{-2})\frac{\dint W}{\dint \bar{I}_2}\,, \nonumber\\
\sigma_2 &=& 2(\lambda^\frac23-\lambda^{-\frac83})\frac{\dint W}{\dint \bar{I}_1} + 2(\lambda^\frac83-\lambda^{-\frac23})\frac{\dint W}{\dint \bar{I}_2} \,.
\label{result_m_13}
\end{eqnarray}
Please note, for a concrete implementation of such a test procedure in a biaxial test rig \cite{ScCaLoKiHoScDeHe17}, for example, on a thin rubber sample as depicted
in Fig.\ \ref{biax-tester}, one can now specify the stretches in the form of displacement boundary conditions for the path of any $m$.
The corresponding stresses then arise and can be measured as force quantities. This allows predictions from the
different material models in the form of energy functions $W$ to be compared with regard to their quality of the stress response in this test procedure.
\clearpage
\subsection{Aspects of {\sc Hill}'s inequality}
\subsubsection{The compressible case}
For a better understanding of the elastic stability concept considered here, it will be useful to summarize the properties of {\sc Hill}'s condition.
Originally proposed by \cite{Hil58,Hil68} in the sense of a rate--condition for compressible response
\begin{equation}
\left<\dfrac{\Dabl^{ZJ}}{\Dabl t}\pb\tau, \f d\right> >0 \quad \forall\,\f d\neq\f 0\,,
\label{Hill_general}
\end{equation}
it can be shown to be equivalent to the following statements \cite{NeHoAgBeSkGhMa25}:
\begin{eqnarray}
&\Longleftrightarrow& \dfrac{\dint\pb\tau}{\dint\log\f v} \quad {\rm is\, a\, positive\, definite\, fourth\, order\, tensor}\nonumber\\
&\Longleftrightarrow& \dfrac{\dint\pb\tau^i(\log\lambda_1, \log\lambda_2,\log\lambda_3)}{\dint\log \lambda_j} \quad {\rm is\, a\, positive\, definite\, symmetric\, matrix}\nonumber \\
&\Longleftrightarrow& W=\widehat{W}(\log \f v)\quad {\rm is\, convex\, in} \log\f v \nonumber\\
&\Longleftrightarrow& W=\widehat{W}(\log\lambda_1, \log\lambda_2,\log\lambda_3)\quad {\rm is\, convex\, in}\, (\log\lambda_1,\log\lambda_2,\log\lambda_3) \nonumber\\
&\Longleftrightarrow& \left<\underbrace{\Dabl^2\widehat{W}(x_1,x_2,x_3)}_{=:\mathbb H}\cdot\eta, \eta\right>_{\mathbb R^3} >0  \quad \forall\,\eta\neq0\in {\mathbb R^3}  \nonumber\\
&\Longleftrightarrow& {\mathbb H}\in{\rm Sym^{++}}(3):\quad {\rm trace}(\mathbb H)>0\,, \quad{\rm trace}({\rm Cof}\, \mathbb H)>0\,, \quad\det\mathbb H>0 \nonumber\\
&\Longleftrightarrow& \left<\pb\tau(\log\f v_1)-\pb\tau(\log\f v_2), \log\f v_1-\log\f v_2\right> >0 \quad \forall\,\f v_1\neq\f v_2\nonumber\\
&\Longleftrightarrow& \sum_{i=1}^3 \left(\pb\tau_i(\log\bar{\lambda}_1,\log\bar{\lambda}_2,\log\bar{\lambda}_3)-\pb\tau_i(\log\lambda_1,\log\lambda_2,\log\lambda_3)\right)(\log\bar{\lambda}_i-\log\lambda_i)>0\nonumber\\
&\Longleftrightarrow& \dint\pb\tau:\dint\pb\epsilon^{\rm log} = \dint\tau_1\,\dint\epsilon_1^{\rm log} + \dint\tau_2\,\dint\epsilon_2^{\rm log} + \dint\tau_3\,\dint\epsilon_3^{\rm log} > 0\,.
\end{eqnarray}
\subsubsection{The incompressible case}
In the incompressible case {\sc Hill's} inequality is evaluated for $\det\f F=1$.
Since
\begin{equation}
\log\det\f F=\log\det\f v={\rm trace}(\log\f v) \,,
\end{equation}
the nonlinear and non--convex constraint $\det\f F=1$ translates with respect to the logarithmic strains $\log\f v$ into the linear and convex constraint
\begin{equation}
0={\rm trace}(\log\f v)=\log\lambda_1+\log\lambda_2+\log\lambda_3 \,.
\end{equation}
Therefore, we have
\begin{eqnarray}
&\Longleftrightarrow& \left<\dfrac{\Dabl^{ZJ}}{\Dabl t}\pb\tau, \f d\right> >0 \quad \forall\,\f d\neq\f 0,\,\, {\rm trace}(\f d)=0 \nonumber\\
&\Longleftrightarrow& W=\widehat{W}(\log\f v) \quad {\rm is\, convex\ in} \log\f v \, {\rm over}\, {\rm trace}(\log\f v)=0 \nonumber\\
&\Longleftrightarrow& W=\widehat{W}(\log\lambda_1,\log\lambda_2,\log\lambda_3) \quad {\rm is\, convex\ in}\nonumber\\
& & \quad\quad\quad\quad\quad\quad (\log\lambda_1,\log\lambda_2,\log\lambda_3)\, {\rm over\,} \log\lambda_1+\log\lambda_2+\log\lambda_3=0 \nonumber\\
&\Longleftrightarrow& \dint\pb\tau:\dint\pb\epsilon^{\rm log} > 0\,\, {\rm with\,\, trace}(\dint\pb\epsilon^{\rm log})=0 \nonumber \\
&\Longleftrightarrow& \dint\tau_1\,\dint\epsilon_1^{\rm log} + \dint\tau_2\,\dint\epsilon_2^{\rm log} + \dint\tau_3\,\dint\epsilon_3^{\rm log} > 0  \quad{\rm with}\quad \dint\epsilon_1^{\rm log} + \dint\epsilon_2^{\rm log} + \dint\epsilon_3^{\rm log} = 0 \nonumber \\
&\Longleftrightarrow& \left<\underbrace{\Dabl^2\widehat{W}(x_1,x_2,x_3)}_{=:\mathbb H}\cdot\eta, \eta\right>_{\mathbb R^3} >0  \quad \forall\,\eta\neq0,\,\, \eta_1+\eta_2+\eta_3=0 \nonumber\\
&\Longleftrightarrow& \left<\underbrace{\f P^{\rm T}\cdot\Dabl^2\widehat{W}(x_1,x_2,x_3)}_{=:\mathbb H^{\rm inc}}\cdot \f P\cdot\xi, \xi\right>_{\mathbb R^2} >0  \quad \forall\,\xi\in\mathbb R^2 \quad {\rm with}\, \f P=\begin{bmatrix} 0 & 1\\ -1 & -1 \\ 1 & 0 \end{bmatrix} \nonumber\\
&\Longleftrightarrow& {\mathbb H^{\rm inc}}\in{\rm Sym^{++}}(2):\quad {\rm trace}(\mathbb H^{\rm inc})>0\,, \quad \det\mathbb H^{\rm inc}>0 \nonumber\\
&\Longleftrightarrow& \mbox{the {\sc Abaqus} check:}\quad{\mathbb D} \,\, {\rm from\, (\ref{D_mat})\, is\,  positive\, definite}\,,
\end{eqnarray}
where we show the last equivalence in Appendix \ref{app_HH_DD}.

\noindent
An evaluation of {\sc Hill's} condition in rate--form (\ref{Hill_general}) along the family
\begin{equation}
\f F={\rm diag}(\lambda,\lambda^m,\lambda^{-(m+1)})
\end{equation}
of the special, i.e.\ "prototype" deformations in Fig.\ \ref{Abaq_Check_InvarEbene} is given in the Appendix~\ref{app-C_Hill}.
A numerical check of {\sc Hill}'s condition for compressibility and a family of elastic energies has recently been given in \cite{GhMaApNe25}.
\section{Derivations and further treatment: the uniaxial case as motivation \label{sec:Deriv}}
The aim of this contribution is a generalized procedure of evaluating (\ref{Drucker_Hill}) for given $W$ in the {\sl plane
of invariants}, see \cite{Tre75} and \cite{BaHoSc12}, as a map of admissible and inadmissible regions for a set of
material parameter $c_1, c_2, c_3, ...$, on which the hyperelastic material description in terms of the elastic energy $W$ depends.

As already known, see \cite{Hol00}, and executed exemplary in e.g.\ \cite{Aba20}, we obtain the stress response (\ref{sig12_HA})
in principal directions and derive here the stability check for the uniaxial case (\ref{tau_1__tau_2__1ax}).

Let us describe the (incremental) stress-strain relationship $\dint \pb \sigma = \mathbb{D} : \dint \pb \epsilon^{\rm log}$ equivalent to (\ref{D_mat}) with
$\dint\pb\epsilon^{\rm log}$ as increment in the {\sc Hencky} strain as in (\ref{Drucker}), i.e.\ the logarithmic strain
$\pb\epsilon^{\rm log} = {\rm diag}(\log\lambda_1,\log\lambda_2,\log\lambda_3)$ by the stretches $\lambda_{1,2,3}$ in principal axes as the main diagonal of a
$[3\times 3]$ tensor/matrix notation, and the fourth order material modulus $\mathbb D$.
In the one--dimensional uniaxial case, the constitutive relation reduces to
\begin{equation}
\dint \sigma =D_{\rm 1D}\, \dint\log\lambda
\label{spg_dehn_1D}
\end{equation}
with $\sigma(\lambda_1,\lambda_2)=\widehat{\sigma}(\log \lambda_1,\log \lambda_2)=\widehat{\sigma}(\lambda,m)=\tau$ for $J=1$, which we refer to in this work.
Following \cite{Ric48} and \cite{WoHoNe25} recently, the {\sc Kirchhoff} stress $\pb \tau$ is conjugate to the logarithmic strain in the sense that
\begin{equation}
\pb\tau = \frac{\dint W(\log \f v)}{\dint\log \f v} \,,
\label{tau_log}
\end{equation}
which is reduced to the one--dimensional case (\ref{tau_1__tau_2__1ax}) considered here.
\nocite{NeHuHoGmBl25}
\nocite{BlHoHuGmNe25}
\begin{figure}[!hbpt]
\centering
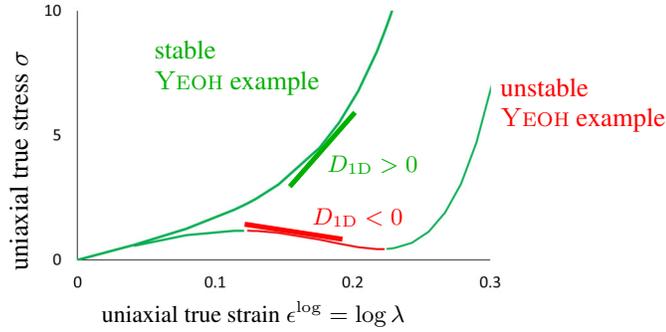
\caption{Example of uniaxial true stress--strain behaviour $\sigma(\log \lambda)$ for the {\sc Yeoh} model (\ref{W_Yeoh}).
In \textcolor{green}{green} a stable and monotone result using the parameters $c_1=1.0$ MPa, $c_2=0.1$ MPa and $c_3=0.3$ MPa;
in \textcolor{red}{red} an unstable and non--monotonic result using the parameters $c_1=1.0$ MPa, $c_2=-0.9$ MPa and $c_3=0.3$ MPa as used below in Tab.\ \ref{tab:YEOHparams}.
The \textcolor{green}{green} and the \textcolor{red}{red} tangents symbolize the slope $D_{\rm 1D}$ of the curves, which is given in (\ref{D_1D_1ax}) below.}
\label{example_monotonicity}
\end{figure}
{Exemplary, we give an uniaxial stress--strain function according to the {\sc Yeoh} model used lateron (\ref{W_Yeoh}) in order to show $D_{1D}$ of (\ref{spg_dehn_1D}) and (\ref{D_1D})$_1$}.

We formulate according to \cite{Aba20}
\begin{equation}
D_{\rm 1D} = \frac{\dint \sigma}{\dint \epsilon^{\rm log}} 
       = \frac{\dint \sigma}{\dint \lambda}\, \frac{\dint \lambda}{\dint\epsilon^{\rm log}}
	   =  \lambda\, \frac{\dint \sigma}{\dint \lambda} \,,
\label{D_1D}
\end{equation}
if we consider the derivative of the natural logarithm
\begin{equation}
\frac{\dint \epsilon^{\rm log}}{\dint \lambda} = \frac{\dint \log \lambda}{\dint \lambda} = \frac{1}{\lambda} \, ,
\end{equation}
and their reversal
\begin{equation}
\frac{\dint \lambda}{\dint \log \lambda} = \lambda
\label{log_eps_inv}
\end{equation}
directly plugged in into (\ref{D_1D})$_2$.

For further investigations we only use (\ref{D_1D})$_3$ and check the expression for monotonicity,
here $D_{\rm 1D}>0$ for a given function $W$, whose stability now depends on the set of
material parameters $c_i$ for any deformation given by $\bar{I}_1$ and $\bar{I}_2$ or $\lambda$ and $m$.
Again, with that formulation (\ref{D_1D})$_3$ enforcing (\ref{Invar_lam_m}), now we are able to come with the representation of the derivatives by
\begin{equation}
\tau=\sigma = \lambda\, \frac{\dint W}{\dint \lambda}
\label{sigma_aus_P}
\end{equation}
as already given by (\ref{tau_1__tau_2__1ax}).
Using (\ref{sigma_aus_P}) to evaluate directly and uniformly for the uniaxial case
\begin{eqnarray}
\label{D_1D_1ax}
D_{\rm 1D} = \lambda\, \frac{\dint\sigma}{\dint\lambda} &=& 4c_1(\lambda^2+\lambda^{-1}) + 8c_2(2\lambda^4-3\lambda^2+\lambda-3\lambda^{-1}+3\lambda^{-2}) \\
                                                                                           &+& 12c_3(3\lambda^6-12\lambda^4+5\lambda^3+9\lambda^2-6\lambda+2+9\lambda^{-1}-18\lambda^{-2}+8\lambda^{-3}) \nonumber
\end{eqnarray}
via (\ref{D_1D})$_3$, we check the material stability by $D_{\rm 1D}\stackrel{?}{>}0$ according to (\ref{Drucker}) and (\ref{Drucker_Hill}), respectively,
and give that example in anticipation of the later used {\sc Yeoh} model (\ref{W_Yeoh}).

Please note that the evaluation of (\ref{D_1D_1ax}) includes the consideration of the derivative with respect to $\log\lambda$ as in (\ref{log_eps_inv}); and the insertion
of the derivative of $W_{\rm Yeoh}$ as the potential for the stresses, which again results in this compact form.
For the general form, see the next section -- also using the {\sc Matlab Symbolic Toolbox} \cite{MATLAB24}.
This settles the question for the uniaxial case.

The case for general $m = (-\frac12 ... 0 ... 1)$ will be shown in the Appendix.
Remember that $D_{1D}>0$ is (only) necessary for {\sc Hill's} inequality (\ref{Drucker_Hill}) to hold \cite{NeHuNgKoMa25} and Appendix. The critical advantage of checking the scalar $D_{1D}>0$ is its
simplicity compared to checking $\mathbb D$ from (\ref{D_mat}) for positive definiteness.
And, in contrast to the {\sc Abaqus} procedure, the check can be easily performed along the whole plane of invariants.
\subsection{Application of {\sc Matlab Symbolic Toolbox} and stability map \label{sec:Matlab_Toolbox}}
We consider here as a first material model the strain energy (density) function according to {\sc Yeoh} \cite{Yeo93} depending only on $\bar{I}_1$
\begin{equation}
W_{\rm Yeoh} = c_1\,(\bar{I}_1-3) + c_2\,(\bar{I}_1-3)^2 + c_3\,(\bar{I}_1-3)^3
\label{W_Yeoh}
\end{equation}
with three parameters $c_1,c_2,c_3$, which is known to possibly satisfy {\sc Hill}'s inequality (\ref{Drucker}) for incompressibility even for negative parameter $c_2$ depending on $c_{1,3}>0$.

Knowing that automatic code generation for mathematical--mechanical implementations has reached a very high level as a result of \cite{kor97} even based on
the newest versions of {\sc Mathematica} \cite{Mathematica24}, we use here, as recently in \cite{FoLePe25}, the {\sc Symbolic Toolbox} of {\sc Matlab} \cite{MATLAB24}
for the implementation of (\ref{W_Yeoh}), (\ref{W_MoRi}), (\ref{W_Ogden}) and (\ref{W_Hencky_quadr_lam_m}),
(\ref{sigma_aus_P}) respectively, and the derivative in (\ref{D_1D})$_3$.
{The {\sc Matlab} code example given here is realized for the {\sc Mooney--Rivlin} strain energy function (\ref{W_MoRi}) in order to show a general case using both
invariants $\bar{I}_1$ and $\bar{I}_2$}

\begin{figure}[!ht]
\centering
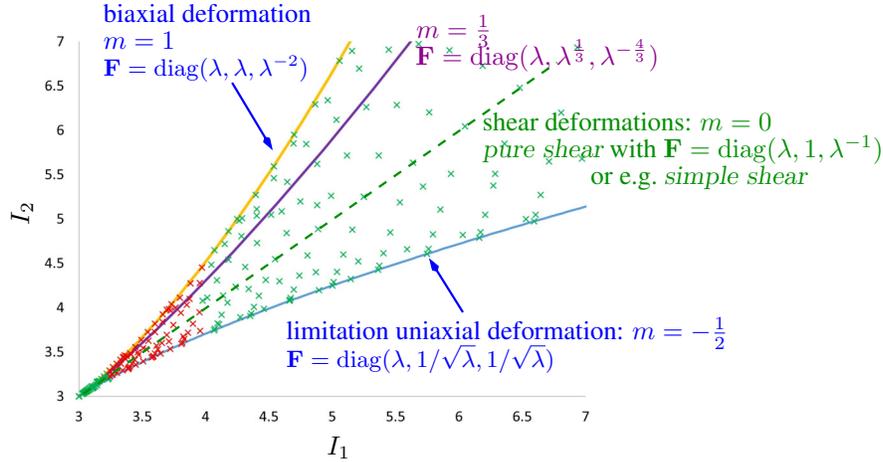 
\caption{Plane of invariants as scope for visualization the material stability of the {\sc Yeoh} model:
The \textcolor{red}{red} crosses here mark the instable region depending on the parameters in Tab.~\ref{tab:YEOHparams}
according to the criterion $D_{1D}^m>0$ of (\ref{D_1D_m}).
In addition, the limiting lines of all possible states given by uniaxial and biaxial deformations are shown. All shear deformation states with $\bar{I}_1\equiv \bar{I}_2$ exactly form the angle bisector of the diagram.
The \textcolor{magenta}{magenta} line visualizes a path for $m=\frac13$ in that plane exemplary.
The density of the points is -- here and in the following figures -- a direct result of the incrementation of $\lambda_\star$ in steps of $\Delta \lambda_\star=0.1$ based on the choosen step size for the stability investigation in {\sc Abaqus} and {\sc Ansys}.
\label{InvarEbene_Yeoh_Stab}}
\end{figure}

\begin{figure}[!htpb]
\centering
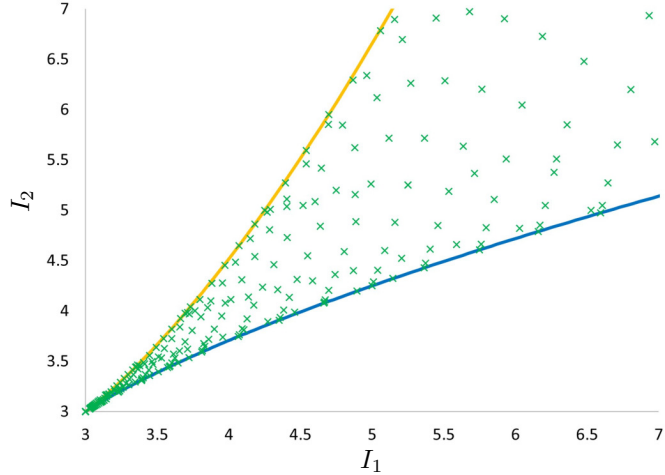
\caption{Plane of invariants as scope for visualization the material stability of the {\sl quadratic} {\sc Hencky} model with material parameter $\mu>0$
according to the criterion $D_{1D}^m>0$ of (\ref{D_1D_m}): The \textcolor{green}{green} crosses here mark the complete stable region. \label{InvarEbene_HENCKY_Stab}}
\end{figure}

\begin{verbatim}
clear all
syms lam m I1 I2 c1 c2

% Mooney-Rivlin
W_MoRi = c1*(I1-3) + c2*(I2-3)

dW_dI1 = diff(W_MoRi,I1);
dW_dI2 = diff(W_MoRi,I2);

%                  I1 = lam^2+lam^(2*m) + lam^(-2*(m+1))
dW_dI1 = subs(dW_dI1,I1,lam^2+lam^(2*m) + lam^(-2*(m+1)))
%                  I2 = lam^(-2)+lam^(-2*m) + lam^(2*(m+1))
dW_dI2 = subs(dW_dI2,I2,lam^(-2)+lam^(-2*m) + lam^(2*(m+1)))

sig1 = 2*(lam^2-lam^(-2*m-2))*dW_dI1 ...
                         + 2*(lam^(2*m+2)-lam^(-2))*dW_dI2
sig1 = collect(sig1,["c1" "c2" "c3"])

sig2 = 2*(lam^(2*m)-lam^(-2*m-2))*dW_dI1 ...
                         + 2*(lam^(2*m+2)-lam^(-2*m))*dW_dI2
sig2 = collect(sig2,["c1" "c2" "c3"])

D_1D_m = diff(sig1,lam)+m*diff(sig2,lam)
D_1D_m = collect(D_1D_m,["c1" "c2" "c3"])

\end{verbatim}
Here, $\tt D\_1D\_m$ is now the complete and analytical representation of (\ref{D_1D}), derived from any strain energy function $W$.
Given $\bar{I}_1$ and $\bar{I}_2$ for any stretch $\lambda$ and mode exponent $m$, we represent $\tt D\_1D\_m$, or (\ref{D_1D}) respectively,
in visualizations as Fig.~\ref{InvarEbene_Yeoh_Stab} -- Fig.\ \ref{InvarEbene_HENCKY_Stab}.  
Here, we give the more general derivation of $D_{1D}$ and $\tau_3=0$ is incorporated in terms of
\begin{equation}
D^m_{1D}=\Dabl_\lambda\widehat\tau_1(\lambda)+m\,\Dabl_\lambda\widehat\tau_2(\lambda)\,,
\label{D_1D_m}
\end{equation}
see (\ref{pos_incr_mod}) in the Appendix \ref{app-C_Hill}, where we use the derivatives
$\Dabl_\lambda\tau_1$ and $\Dabl_\lambda\tau_2$ of the principal stress components (\ref{tau_1__tau_2}) obtained by the help of the {\sc Matlab Symbolic Toolbox} \cite{MATLAB24},
shown in the last two lines of the code example above.

\begin{table}[!htpb]
\caption{Exemplary material parameters for the {\sc Yeoh} model in MPa = $\frac{\rm N}{{\rm mm}^2}$.}
\label{tab:YEOHparams} 
\begin{tabular}{c|c|c}
$c_1$ & $c_2$ & $c_3$  \\
\noalign{\smallskip}\hline\noalign{\smallskip}
1.0 & -0.9 & 0.3
\end{tabular}
\end{table}

\begin{figure}[!ht]
\centering
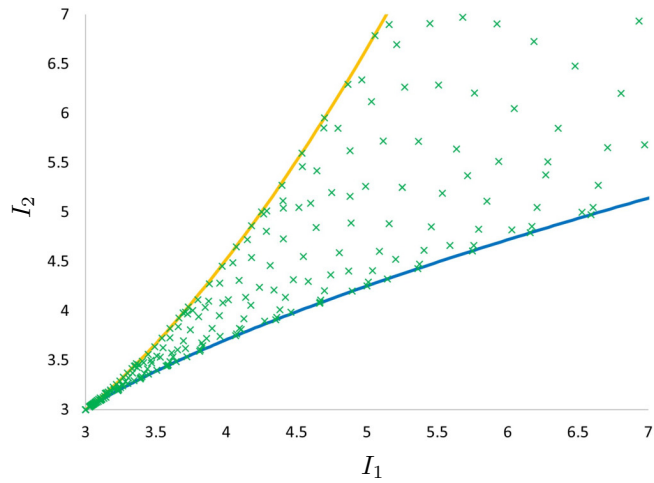
\caption{Plane of invariants as scope for visualization of the material stability of the {\sc Mooney--Rivlin} model according to the criterion $D_{1D}^m>0$:
The \textcolor{green}{green} crosses mark the completely stable region depending on the parameters in Tab.\ \ref{tab:MoRiparams_1}.}
\label{InvarEbene_MoRi_Stab_1}
\end{figure}

\begin{figure}[!ht]
\centering
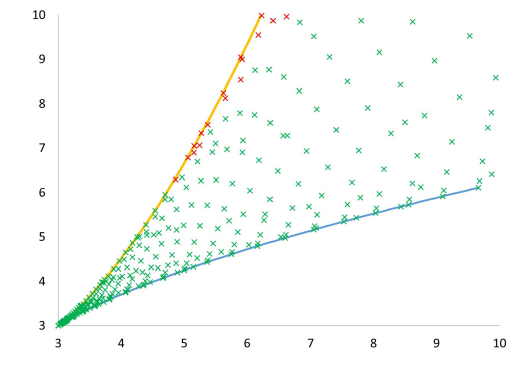
\caption{Plane of invariants as scope for visualization the material stability of the {\sc Mooney--Rivlin} model:
The \textcolor{red}{red} crosses here mark the instable region depending on the parameters in Tab.\ \ref{tab:MoRiparams_2}
according to the criterion $D_{1D}^m>0$ of (\ref{D_1D_m}): $c_1=0.8$ MPa and $c_2=-0.2$ MPa}.
\label{InvarEbene_MoRi_Stab_2}
\end{figure}

Such visualization in the form of a stability map in the plane of invariants enables a rapid identification of unstable
parameter ranges and thus provides a tool for targeted parameter adjustment during material calibration.
Thus, we are now able to show by the proposed procedure visualized in a "heat" map,
where \textcolor{green}{\sl green} indicates the possibly stable region, whereas 
\textcolor{red}{\sl red} the instable zone due to the parameter declaration for given deformations.
{Please note, that any shear deformation states (both {\sl plane strain}, i.e.\ {\sl pure shear}, with principal axes are fixed and {\sl simple shear} with rotating principal axes)
with $\bar{I}_1=\bar{I}_2$ exactly form the angle bisector of the diagram.}
In this first exemplary case with parameters from Tab.\ \ref{tab:YEOHparams}, a picture emerges as in Fig.\ \ref{InvarEbene_Yeoh_Stab}, {where we provoked the instability by the negative parameter $c_2<0$}.
Here, it is clear that after an initially stable range near to the stress free origin, i.e.\ $\bar{I}_1=\bar{I}_2=3$, the material behavior becomes unstable across the range of all possible deformations, see red crosses.
Using the procedure presented here, it is now very easy to adjust, for example, the parameter $c_2$ to eliminate instability.
Of course, the resulting adjustment must also be consistent with the experimental procedures performed on the material or components.
As a further example, we demonstrate the application to the incompressible {\sc Mooney-Rivlin} model \cite{RiSa51}, known as the simplest model with $I_1$ and $I_2$ dependence:
\begin{equation}
W_{\rm MoRi} = c_1\,(I_1-3) + c_2\,(I_2-3) \,.
\label{W_MoRi}
\end{equation}
This model is known for its supposedly good fit to uniaxial tension experiments.
{Firstly, we show the results of our approach for positive parameters as given in Tab.\ \ref{tab:MoRiparams_1}, which satisfy {\sc Hill}'s condition everywhere and
consistently yield a picture of stable results as in  Fig.\ \ref{InvarEbene_MoRi_Stab_1}.}
\begin{table}[!ht]
\caption{Exemplary material parameters for the {\sc Mooney--Rivlin} model in MPa = $\frac{\rm N}{{\rm mm}^2}$.}
\label{tab:MoRiparams_1} 
\begin{tabular}{c|c}
$c_1$ & $c_2$  \\
\noalign{\smallskip}\hline\noalign{\smallskip}
0.8 & +0.6
\end{tabular}
\end{table}
Secondly, one can show a counterexample here: unfortunately, this ignores the fact that such a fit, here with negative parameter $c_2$,
can lead to unstable behavior in regions of shear and biaxial tension. This is highlighted again by the red markers in Fig.\ \ref{InvarEbene_MoRi_Stab_2}.
\begin{table}[!ht]
\caption{Exemplary material parameters for the {\sc Mooney--Rivlin} model in MPa = $\frac{\rm N}{{\rm mm}^2}$.}
\label{tab:MoRiparams_2} 
\begin{tabular}{c|c}
$c_1$ & $c_2$  \\
\noalign{\smallskip}\hline\noalign{\smallskip}
0.8 & -0.2
\end{tabular}
\end{table}
Such parameter combinations often lead to non--convergent or unrealistic behavior in FEM simulations and should
therefore be excluded in advance by stability analysis, i.e.\ even though a supposedly good fit was achieved in the uniaxial tension.
In Appendix \ref{app-MoRi_ext} we reconsider $W_{\rm MoRi}$ (\ref{W_MoRi}) in order to have a more
detailed look into stability statements incorporating the two invariants $I_1$ and $I_2$.

As a next example, we describe a somewhat more complicated case, in that the derivation of this procedure is applied to the {\sc Ogden} model(s) \cite{Ogd72},
i.e. a formulation in the stretches and not directly in the invariants,  as described in detail in \cite{HoMu25} for special cases:
Here, we consider exemplary the case of a 2nd order model
\begin{equation}
W_{\rm Ogden} = \frac{2\mu_1}{\alpha_1^2}(\lambda_1^{\alpha_1}+\lambda_2^{\alpha_1}+\lambda_3^{\alpha_1}-3)+\frac{2\mu_2}{\alpha_2^2}(\lambda_1^{\alpha_2}+\lambda_2^{\alpha_2}+\lambda_3^{\alpha_2}-3)
\label{W_Ogden}
\end{equation}
with fixed $\alpha_1:=6$ and $\alpha_2:=-4$, which is then equivalent to
\begin{equation}
W_{\rm Ogden} = \frac{\mu_1}{18}(\bar{I}_1^3-3\bar{I}_1\bar{I}_2) + \frac{\mu_2}{8}(\bar{I}_2^2-2\bar{I}_1-3) \,,
\label{W_Ogden_Inv}
\end{equation}
following \cite{HoMu25} as advanced combination of (2.10)$_2$ and (2.12)$_2$ therein.
Further combinations of this model class are discussed in \cite{HoMu25}; a generalization using the closed--form conversion of a deformation state in the form of the stretches into the invariants given there, is still pending for the application in this work, see Appendix \ref{app-A}, without diminishing the achievements presented here.
\begin{table}[!htpb]
\caption{Exemplary material parameters for the {\sc Ogden} model in MPa = $\frac{\rm N}{{\rm mm}^2}$.}
\label{tab:OGDENparams} 
\begin{tabular}{c|c}
$\mu_1$ & $\mu_2$  \\
\noalign{\smallskip}\hline\noalign{\smallskip}
0.3 & -0.2
\end{tabular}
\end{table}
\begin{figure}[!htpb]
\centering
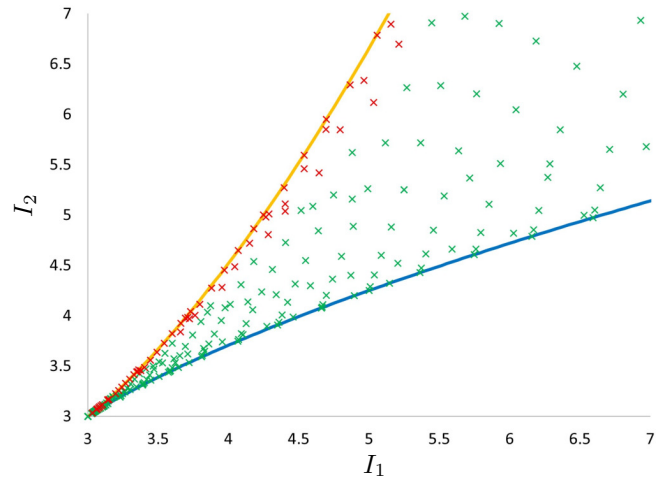
\caption{Plane of invariants as scope for visualization the material stability of the {\sc Ogden} model:
The \textcolor{red}{red} crosses here mark the instable region depending on the parameters in Tab.~\ref{tab:OGDENparams}
according to the criterion $D_{1D}^m>0$ of (\ref{D_1D_m}): $\mu_1=0.3$ MPa and $\mu_2=-0.2$ MPa.}
\label{InvarEbene_OGDEN_Stab}
\end{figure}
However, this example, as before, shows that a supposedly good fit to uniaxial tension data causes unstable behavior in other deformation modes. This must be avoided in simulations.

{As a final and more extensive example, we apply the stability analysis procedure presented here to the {\sl quadratic} {\sc Hencky} {\sl model} proposed in \cite{NeEiMa15}, which is given by
\begin{equation}
W_{\rm Hencky} = \mu\, \lVert\log\f v\rVert^2 + \frac{\kappa}{2}\, {\rm tr}^2(\log \f v)
\label{W_Hencky_quadr_star}
\end{equation}
with $\mu>0$ as shear modulus and $\kappa>0$ as bulk modulus and ${\rm tr}(.)$ as matrix {\sl trace} operator.
Please note that this model in the {\sl quadratic} form with "$\log\f v$" as strain tensor is not yet available in e.g.\ {\sc Abaqus} or other commercial FE--codes.}

{In the incompressible case with $\lambda_3=(\lambda_1\,\lambda_2)^{-1}$ this model (\ref{W_Hencky_quadr_star}) reduces to
\begin{equation}
W_{\rm Hencky} = 2\mu\, (\log^2 \lambda_1 + \log^2 \lambda_2 + \log \lambda_1\, \log \lambda_2) \,.
\label{W_Hencky_quadr}
\end{equation}
Applying here the same arguments for the prototype deformations as in (\ref{Invar_lam_m}) from above, (\ref{W_Hencky_quadr}) is directly reformulated into
\begin{equation}
W_{\rm Hencky} = 2\mu\, \left(\frac{2+m}{\lambda}\,\log\lambda + \frac{2m+1}{\lambda}\,\log(\lambda^m)\right)
\label{W_Hencky_quadr_lam_m}
\end{equation}
in terms of the {\sl stretch intensity} $\lambda$ and the {\sl mode parameter} $m$.
The stress computation (\ref{tau_1__tau_2}) allows us to reuse (\ref{W_Hencky_quadr_lam_m}) directly in (\ref{D_1D}), which results in Fig.~\ref{InvarEbene_HENCKY_Stab}
 for $\mu=1$ MPa showing just stable situations in the plane of invariants.}

Since
%
$\pb\tau=\dfrac{\dint W_{\rm Hencky}}{\dint \log\f v}=2\mu\, \log\f v$
%
we see $\pb\tau$ to clearly satisfy the {\sc Hill} mono\-to\-nicity condition (\ref{Drucker_Hill}).
However, the shear stress component $\tau_{12}=2\mu\, (\log\f v)_{12}$ with $\log\f v=\frac12\log\f b$ is non--monotone in $\gamma$, as we show in the Appendix \ref{app-B_SiSh}.
Again we see that cases with rotating principal axes are not detected as unstable by the procedure above and additional criteria must be applied
as suggested in \cite{WoHoNe25} and \cite{NeHoAgBeSkGhMa25}:
namely to additionally require for stability the "LH--ellipticity" of the constitutive law.

{In contrast, a nonlinear extension of the model (\ref{W_Hencky_quadr_star}) is the {\sl exponentiated} {\sc Hencky} model
\begin{equation}
W_{\rm expH}=\mu\, \exp(||\log\f v||^2) + \frac{\lambda}{2}\, \exp((\log\det\f F)^2)
\end{equation}
see \cite{NeGhLa15} and \cite{NeMaNeBa18} for a FE implementation.
This model shows a monotonic behavior in the 12--true--shear--stress components.}
%
%
%
%
%
\subsection{Comparison with the {\sc Abaqus} stability check of $\mathbb D\in{\rm Sym}^{++}(2)$ \label{sec:ABAQ_check}}
We refer to Section \ref{sec:Background}, where we discuss in detail the basic approach of the stability check of {\sc Abaqus} as presented in \cite{Aba20},
section "Hyperelastic behavior of rubberlike materials", and present our extension to the entire plane of invariants.
\begin{figure}[!ht]
\centering
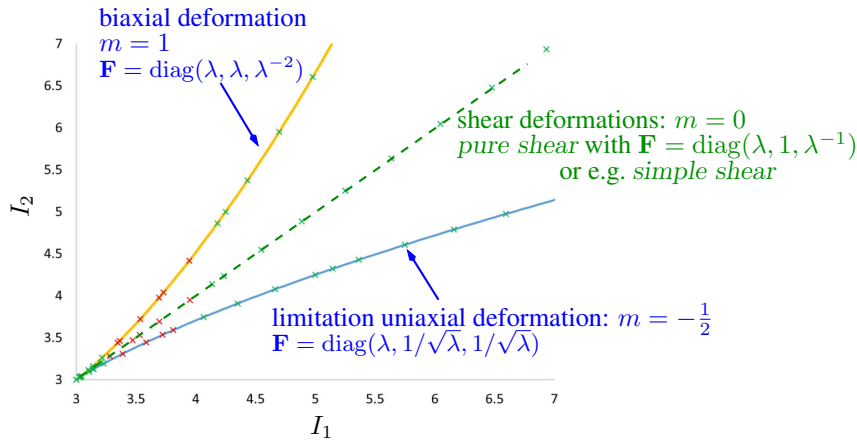 
\caption{Plane of invariants as scope for visualization of the {\sc Abaqus} stability check for the {\sc Yeoh} model with the parameters in Tab.\ \ref{tab:YEOHparams}.
\label{InvarEbene_Yeoh_Stab__ABAQ}}
\end{figure}
This figure (Fig.\ \ref{InvarEbene_Yeoh_Stab__ABAQ}) should be compared to Fig.\ \ref{InvarEbene_Yeoh_Stab} showing the material stability situation for the entire region of incompressible deformation,
whereas here in Fig.\ \ref{InvarEbene_Yeoh_Stab__ABAQ} the stability check is just performed along the three prototype deformations by evaluating
(\ref{tau_1__tau_2__1ax}) -- (\ref{tau_1__tau_2__biax}) mentioned in the {\sc Abaqus} {\sl Manual} \cite{Aba20}.

In contrast and extension to that, here we give the results by exploring (\ref{tau_1__tau_2}) and thus apply the evaluation of (\ref{D_mat}) for the whole region.
\begin{figure}[!ht]
\centering
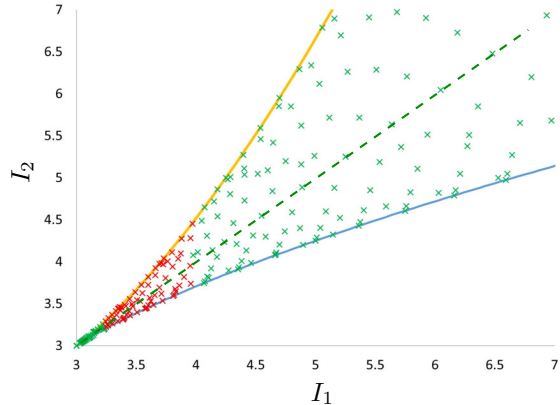 
\caption{Plane of invariants as scope for visualization of the stability check using $\mathbb D\in{\rm Sym}^{++}(2)$ of (\ref{D_mat}) in the entire region for the {\sc Yeoh} model with the parameters in Tab.\ \ref{tab:YEOHparams}.
\label{InvarEbene_Yeoh_Stab__ABAQ_ganzeEbene}}
\end{figure}
As can be seen, all three methods end up in the exactly same results -- here for the {\sc Yeoh} model.

%
%
%
%
\section{Conclusions}
We have succeeded in developing a uniform representation and a standardized procedure that makes it possible to visualize material stability for the incompressible isotropic hyperelasticity model class and its prototype deformation modes in an overall picture.
A similar procedure (presumably numerically) is also used in commercial FE--codes such as {\sc Abaqus} or {\sc Ansys} to warn of potential instabilities during application. This is now also possible in a comprehensible and graphical way,
and should definitely be included when adapting material properties to material samples or component measurements for industrial applications.
It must be noted that deformation modes with rotating principal axes -- the simple shear case is a good example -- can fulfill the incompressible {\sc Hill} stability condition $\mathbb H^{\rm inc}\in {\rm Sym}^{++}(2)$,
but this does not guarantee monotonicity of the true shear stress--strain relationship in
{\sl simple shear}\footnote{We show in the Appendix \ref{app-B_SiSh} the relation between the shear component and the shear $\gamma$.}, see \cite{NeHuNgKoMa25}, and
 therefore one should in addition require as useful stability condition the "{\sc Legendre--Hadamard}--ellipticity" of the constitutive law, namely the condition
$\Dabl^2_{\f F}W(\f F).(\xi\otimes\eta,\xi\otimes\eta)>0 \quad \forall\, \xi,\eta\ne 0$.
This computation of {\sc Hill}'s stability requirement and the additional LH--ellipticity seems to be sufficient for excluding any sort of material instability in
incompressible hyperelasticity and should therefore be routinely checked for FE--calculation in accordance with the findings in \cite{NeHuNgKoMa25}. It remains to find a structure that a priori satisfies {\sc Hill's} inequality {\bf and} the {\sc Legendre--Hadamard} ellipticity while being able to suitably fit experimental data. This will be pursued next in \cite{KlWoBaNe25} using a neural network structure.
%
%
%
%
%
\section*{Conflict of interest}
The author declares that he has no conflict of interest.
\section*{Acknowledgement}
The author thanks {\sc Patrizio Neff} (Faculty of Mathematics, Univ.\ Duisburg--Essen, Ger\-many) for in--depth discussions of pertinent stability concepts in incompressible isotropic hyperelasticity. Especially,
the essential equivalence of $\mathbb H^{\rm inc}$ and $\mathbb D$ as stability indicators in the Appendix \ref{app_HH_DD} is an outcome of this discussion, together with Appendixes
\ref{app-C_Hill}, \ref{app-MoRi_ext} and \ref{app-B_SiSh}.

The author also thanks {\sc Attila Kossa} ({\sl Department of Applied Mechanics},
Budapest University of Technology and Economics) for sharing a detailed numerical discussion on the stability limits of the {\sc Mooney--Rivlin} model.
\begin{appendix}
\section{Derivations within the plane of invariants}
\subsection{A closed form representation of stretches versus invariants at incompressibility \label{app-A}}
In \cite{HoMu25} the closed form representation of \cite{YaChHu22} is reformulated and given in order to obtain the stretches ($\lambda_1, \lambda_2$) from a given
pair ($\bar{I}_1, \bar{I}_2$) for the incompressible case, i.e.\ $I_3\equiv 1$.
In a similar way, mathematically somewhat more complex, but also in closed form, one can obtain the representation \cite{BaHoSc12} used here by ($\lambda, m$) from such a given pairing ($\lambda_1, \lambda_2$) or ($\bar{I}_1, \bar{I}_2$), respectively.
This will be demonstrated here as an example.
Let us start with two substitutions in order to simplify our notation: We set
\begin{equation}
\mu:=\lambda^2 \quad {\rm and} \quad \eta:=\mu^m
\end{equation}
to obtain the given invariants ($\bar{I}_1, \bar{I}_2$) from (\ref{Invar_lam_m}) as
\begin{equation}
\label{Invar_mu_eta}
\bar{I}_1=\mu+\eta + \frac{1}{\mu\, \eta} \quad {\rm and} \quad \bar{I}_2=\frac{1}{\mu}+\frac{1}{\eta}+\mu\, \eta \,.
\end{equation}
Here, (\ref{Invar_mu_eta})$_1$ represents the quadratic equation $\mu\eta^2+(\mu^2-\bar{I}_1\mu)\eta+1=0$ for $\eta$, whose solution is given by
\begin{equation}
\eta_{1,2}=\frac{(\bar{I}_1\mu-\mu^2)\pm\sqrt{\mu^4-2\bar{I}_1\mu^3+\bar{I}_1^2\mu^2-4\mu\,}}{2\mu} = \mu^m \,.
\label{lsg_eta}
\end{equation}
%
On the other hand, (\ref{Invar_mu_eta})$_2$ provides the cubic representation $\mu^3-\bar{I}_1\mu^2+\bar{I}_2\mu-1=0$ in $\mu$, which can be transformed by
\begin{equation}
p:=+\bar{I}_2-\frac{\bar{I}_1^2}{3} \quad {\rm and} \quad q:=+\frac{\bar{I}_1\, \bar{I}_2}{3}-\frac{2}{27}\bar{I}_1^3-1
\end{equation}
into the {\sl reduced cubic equation} $z^3+pz+q=0$ with $z:=\mu-\frac{\bar{I}_1}{3}$.
In our case, this equation has three real solutions, where we bypass the complex solutions using $\rho:=\sqrt{-p^3/27}$ and $\cos \varphi=-q/2\rho$.
This gives us $z_1=2\sqrt[3]{\rho}\,\cos(\varphi/3)$ as the first of the three solutions, from which $\mu=\lambda^2$ follows directly.

With $\eta^+=\mu^m$ from (\ref{lsg_eta})$_1$ we have
\begin{equation}
\log \eta^+ = m\, \log \mu
\end{equation}
and therefore $m=\frac{\log \eta^+}{\log \mu}$ from that derivations.
This allows us to convert any representation into a correspondingly different one, even if it involves nonlinear equations up to the 3rd order.
\subsection{Analytical derivation of boundary lines in the plane of invariants \label{app-B_boundarylines}}
We give the analytical derivation of the functions $\bar{I}_2=\bar{I}_2(\bar{I}_1)$ representing the boundary lines of possible deformation modes in the plane of invariants, see Fig.\ \ref{Abaq_Check_InvarEbene}.
The two boundary lines for the uniaxial and the biaxial deformation mode, respectively, are obtained using (\ref{Invar_1_2}).

\noindent
From (\ref{Invar_1_2})$_1$ with $\lambda_1=\lambda$ and $\lambda_2=\lambda^{-\frac12}$ we see
\begin{equation}
\lambda^3-\bar{I}_1\lambda+2=0,
\end{equation}
which can be solved for $\lambda$ directly applying {\sc Cardano's} formula.
Assigning $p:=-\bar{I}_1$, $q:=2$ and therefore $D=1-\bar{I}_1^3/27<0$, we receive
\begin{equation}
\varphi=\frac13\arccos\left[-\sqrt{27/\bar{I}_1^3\,\,}\right]
\label{phi_uniax}
\end{equation}
and from this the stretch
\begin{equation}
\lambda = 2\,\sqrt{\frac{\bar{I}_1}{3}\,\,}\cos\varphi\,.
\end{equation}
Using that result in (\ref{Invar_1_2})$_2$ we obtain
\begin{equation}
I_2=4\,\sqrt{\frac{\bar{I}_1}{3}\,\,}\cos\varphi+\frac{3}{4\, \bar{I}_1\cos^2\varphi} \,,
\label{I2_I1_line_1ax}
\end{equation}
which represents (with $\varphi$ from (\ref{phi_uniax})) the uniaxial boundary line $\bar{I}_2(\bar{I}_1)$ and has already been published in \cite{BaHoSc12}.
The boundary line for the biaxial mode is obtained in the same way from (\ref{Invar_1_2}) with $\lambda_1=\lambda_2=\lambda$.

In addition, we again give the {\sl plane of invariants} with the boundary lines (\ref{I2_I1_line_1ax}) from above and points of data pairs $(\bar{I}_1, \bar{I}_2)$ in \textcolor{blue}{blue}
representing a set of possible deformations generated by a sequence of randomly produced deformation gradients $\f F$ with $J=\det\f F\equiv 1$.
\begin{figure}[!htpb]
\centering
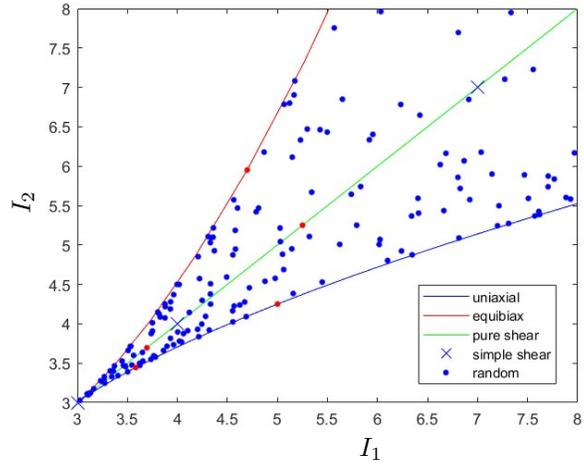
\caption{Plot and derivation of {\sl plane of invariants} with randomly generated points of possible incompressible deformations in \textcolor{blue}{blue}.}
\label{invar_random}
\end{figure}

\begin{figure}[!htpb]
\centering
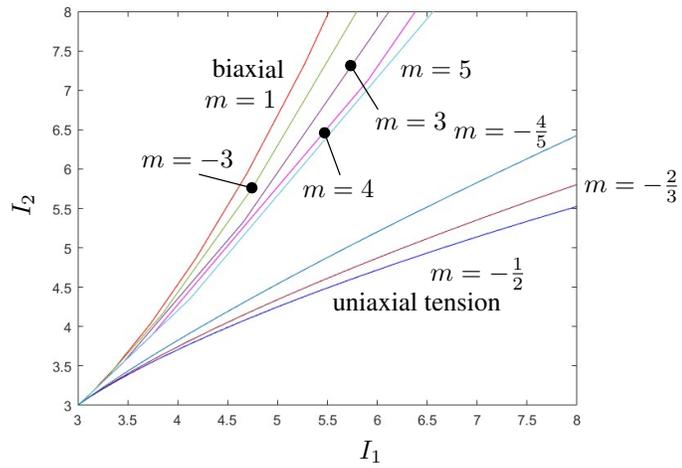
\caption{Plot of deformation paths for arbitrary $m$ in the plane of invariants.}
\label{invar_m4}
\end{figure}

Furthermore, one might ask what happens if one chooses values of $m$ in the family $\f F={\rm diag}(\lambda,\lambda^m,\lambda^{-(m+1)})$ that lie outside of the interval $m \in (-\frac12,1)$.
Here, we show the path of different $m$ in the plane of invariants as an example, see Fig.\ \ref{invar_m4}.

If we now consider a possible deformation path within that family, we see that such situations can be interpreted as a sequence of successive prototype deformations
\begin{equation}
\f F_{(m=4)}=\begin{bmatrix} \lambda \quad 0 \quad 0 \\
                                                0 \quad \lambda^4 \quad 0 \\
                                               0 \quad 0 \quad \lambda^{-5} \end{bmatrix}
=\underbrace{\begin{bmatrix} \lambda \quad 0 \quad 0 \\
                                                0 \quad \lambda \quad 0 \\
                                               0 \quad 0 \quad \lambda^{-2} \end{bmatrix}}_{\rm biax}\cdot
\underbrace{\begin{bmatrix} 1 \quad 0 \quad 0 \\
                                                0 \quad \lambda^3 \quad 0 \\
                                               0 \quad 0 \quad \lambda^{-3} \end{bmatrix}}_{\rm multiple\, pure\, shear}
\label{F_sequence}
\end{equation}
by a multiplicative decomposition of $\f F$, see \cite{Lee69} or \cite{HuBa03}.

Here, we again refer to Fig.\ \ref{m_equiv}, where we have shown the equivalence of $m$ and $m^\star$ in tension and compression, i.e.\ $\lambda>1$ and
$\lambda<1$, and vice versa.
Reformulating (\ref{Invar_mu_eta}) from above, we obtain a quadratic equation in $\mu$ with the solutions $\mu_{1,2}$ and the cubic equation
\begin{equation}
\label{invars_eta}
\eta^3-I_1\eta^2+I_2\eta-1 = 0
\end{equation}
in $\eta=\mu^m$ with the at first three possible solutions $\eta_{1,2,3}$.
In consequence, we have a set of six possible solution of (\ref{invars_eta}), i.e.\ $\eta_{1,2,3}=\mu_{1,2}^{m_{1,2,3}}$.
Unfortunately, we cannot yet provide a further analytical solution. Therefore, we are currently reliant on a numerical solution and a comparison of possible meaningful results.

This comparison yields only one meaningful $m^\star$ for $\lambda\in (0.1,1)$ and a given, fixed mode parameter $m$, where we exclude $m^\star=m$ and $m^\star=0$ as solutions. At this point, we define the here resulting $m^\star$ as equivalent to the given $m$ as depicted in Fig.\ \ref{m_equiv}.
\section{{\sc Matlab} code for the stress computation \label{app-C_ABAQ_stabi}}
Here we give the  {\sc Matlab} code to calculate the two stress components from (\ref{sig12_HA}) in order to reproduce the ABAQUS stability criterion according to (\ref{D_mat}):
\begin{verbatim}
clear all
syms lam1 lam2 lam3 I1 I2 c1 c2 c3

W_Yeoh = c1*(I1-3) + c2*(I1-3)^2 + c3*(I1-3)^3
dW_dI1 = diff(W_Yeoh,I1);
dW_dI2 = diff(W_Yeoh,I2);

%                  I1 = lam1^2+lam2^2+lam3^2
dW_dI1 = subs(dW_dI1,I1,lam1^2+lam2^2+lam3^2);
%                  I2 = lam1^2*lam2^2+lam1^2*lam3^2+lam2^2*lam3^2
dW_dI2 = subs(dW_dI2,I2,lam1^2*lam2^2+lam1^2*lam3^2+lam2^2*lam3^2);

% sig3 == 0
p = 2*lam3^2*dW_dI1 + 2*(lam1^2*lam3^2+lam2^2*lam3^2)*dW_dI2;

sig1 = lam1*2*lam1*dW_dI1 + lam1*(2*lam1*lam2^2+2*lam1*lam3^2)*dW_dI2 - p;
sig2 = lam2*2*lam2*dW_dI1 + lam2*(2*lam1^2*lam2+2*lam2*lam3^2)*dW_dI2 - p;

% Incompressibility: lam3 = 1/lam1/lam2
sig1 = subs(sig1,lam3,1/lam1/lam2)
sig2 = subs(sig2,lam3,1/lam1/lam2)

% ABAQ - multiplication by lam_i is equiv. to D_log(.) 
DD_11 = simplify( expand( lam1*diff(sig1,lam1) ) );
DD_12 = simplify( expand( lam2*diff(sig1,lam2) ) );
DD_21 = simplify( expand( lam1*diff(sig2,lam1) ) );
DD_22 = simplify( expand( lam2*diff(sig2,lam2) ) );

det = DD_11*DD_22 - DD_12*DD_21
trc = DD_11+DD_22

% uniax
det_1ax = subs(det,lam2,lam1^(-1/2))
trc_1ax = subs(trc,lam2,lam1^(-1/2))

% PS
det_PS = subs(det,lam2,1)
trc_PS = subs(trc,lam2,1)

% biax
det_biax = subs(det,lam2,lam1)
trc_biax = subs(trc,lam2,lam1)
\end{verbatim}
Getting these expressions in analytical form, we are able to plot the results depending on the parameters $c_1, c_2, c_3$.
\section{Evaluation of {\sc Hill's} condition in rate--form \label{app-C_Hill}}
In continuation of (\ref{Hill_general}) in rate--form adapting a calculation from \cite{NeHuNgKoMa25}, Sect.\ 4.2,
{\sc Hill's} inequality for diagonal deformation states amounts to
\begin{equation}
0<\left<\dfrac{\Dabl^{ZJ} \pb\tau}{\Dabl t}, \f d\right>=\left<\dfrac{\partial \pb\tau}{\partial t}, \f d\right>
=\sum_{i=1}^3\, \partial_t\left[\tau_i\left(\lambda_1(t),\lambda_2(t),\lambda_3(t)\right)\right]\, \frac{\dot{\lambda}_i(t)}{\lambda_i(t)}\,,
\label{Hill_Appendix}
\end{equation}
where we obtain the rate $\f d={\rm sym}\, \f l=\frac12(\f l+\f l^{\rm T})$ from $\f l=\dot{\f F}\cdot \f F^{-1}$ here with
\begin{equation}
\f F={\rm diag}\,(\lambda(t),\lambda^m(t),\lambda^{-m-1}(t))
\end{equation}
and $\det \f F=1$ resulting in
\begin{equation}
\dot{\f F}={\rm diag}\left(\dot{\lambda}(t), m\, \lambda^{m-1}(t)\,\dot{\lambda}(t),(-m-1)\,\lambda^{-m-2}\,\dot{\lambda}(t)\right)
\end{equation}
and therefore
\begin{eqnarray}
\f d &=&{\rm diag}\left(\frac{\dot{\lambda}(t)}{\lambda(t)}, m\, \frac{\lambda^{m-1}}{\lambda^m}\,\dot{\lambda}(t),(-m-1)\,\frac{\lambda^{-m-2}}{\lambda^{-m-1}}\,\dot{\lambda}(t)\right)\nonumber\\
       &=&{\rm diag}\left(\frac{\dot{\lambda}(t)}{\lambda(t)}, m\, \frac{\dot{\lambda}(t)}{\lambda(t)},(-m-1)\,\frac{\dot{\lambda}(t)}{\lambda(t)}\right)\nonumber\\
       &=&\frac{\dot{\lambda}(t)}{\lambda(t)}\,{\rm diag}(1, m,-(m+1)) \quad{\rm with}\,\, {\rm trace}(\f d)=0\,.
\end{eqnarray}
\noindent
Going on, (\ref{Hill_Appendix}) writes now as
\begin{eqnarray}
0&<&\left<\dfrac{\partial \pb\tau}{\partial t}, \f d\right>=\dfrac{\partial \pb\tau}{\partial t}:\f d=\partial_t\left[\tau_1(.)\right]\,\frac{\dot{\lambda}(t)}{\lambda(t)}
                                                                                        +\partial_t \left[\tau_2(.)\right]\,m\, \frac{\dot{\lambda}(t)}{\lambda(t)}
                                                                                        -\partial_t \left[\tau_3(.)\right]\, (m+1)\,\frac{\dot{\lambda}(t)}{\lambda(t)}\nonumber\\
&=&\left(\partial_t \left[\tau_1(.)\right]+m\,\partial_t \left[\tau_2(.)\right]-(m+1)\, \partial_t \left[\tau_3(.)\right] \right)\,\frac{\dot{\lambda}}{\lambda} \\
&=&\left(\partial_t \left[\tau_1(\lambda,\lambda^m,\lambda^{-m-1})\right]
             +\partial_t \left[\tau_2(\lambda,\lambda^m,\lambda^{-m-1})\right]
             +\partial_t \left[\tau_3(\lambda,\lambda^m,\lambda^{-m-1})\right]\right)\,\frac{\dot{\lambda}}{\lambda}\nonumber\,.
\end{eqnarray}
We define
\begin{eqnarray}
\widehat\tau_1(\lambda(t)) &:=& \tau_1(\lambda(t),\lambda^m(t),\lambda^{-m-1}(t)), \nonumber\\
\widehat\tau_2(\lambda(t)) &:=& \tau_2(\lambda(t),\lambda^m(t),\lambda^{-m-1}(t)), \\
\widehat\tau_3(\lambda(t)) &:=& \tau_3(\lambda(t),\lambda^m(t),\lambda^{-m-1}(t)) \nonumber
\end{eqnarray}
and observe that $\partial_t\left[\widehat{\pb\tau}_i\left(\lambda(t)\right)\right]=\Dabl_\lambda\widehat{\pb\tau}_i(\lambda(t))\, \dot{\lambda}(t)$ continuing further
\begin{eqnarray}
0<\left<\dfrac{\partial \pb\tau}{\partial t}, \f d\right>&=&\left\{\Dabl_\lambda \widehat\tau_1(\lambda)\,\dot{\lambda}+m\,\Dabl_\lambda\widehat\tau_2(\lambda)\,\dot{\lambda}-(m+1)\,\Dabl_\lambda\widehat\tau_3(\lambda)\,\dot{\lambda}\right\}\,\frac{\dot{\lambda}}{\lambda}\nonumber\\
&=&\left\{\Dabl_\lambda\widehat\tau_1(\lambda)+m\,\Dabl_\lambda\widehat\tau_2(\lambda)-(m+1)\,\Dabl_\lambda{\widehat\tau_3(\lambda)}\right\}\underbrace{\frac{\dot{\lambda}\,\dot{\lambda}}{\lambda}}_{>0}\,.
\label{Hill_2D}
\end{eqnarray}
%
%
Let us check the three prototype modes of deformation
\begin{table}[!h]
\begin{tabular}{llll}
uniaxial tension & with $m=-\frac12$, & $\f F={\rm diag}(\lambda,\lambda^{-\frac12},\lambda^{-\frac12})$ & and $\tau_2=\tau_3=0$ \\
pure shear & with $m=0,$ & $\f F={\rm diag}(\lambda,1,\lambda^{-1})$ &and $\tau_3=0$ \\
$\Longleftrightarrow$ {\sl plane strain}  & with $\lambda_2=1$ fixed \\
$\Longleftrightarrow$ {\sl planar tension} \\
biaxial tension & with $m=1$, & $\f F={\rm diag}(\lambda,\lambda,\lambda^{-2})$ & and $\tau_1=\tau_2, \,\,\tau_3=0$\,.
\end{tabular}
\end{table}

\noindent
In all three cases, fully considering the stress--boundary conditions, we obtain 
\begin{equation}
0<\left<\dfrac{\partial \pb\tau}{\partial t}, \f d\right> \Longleftrightarrow
\underbrace{\Dabl_\lambda\widehat\tau_1(\lambda)}_{\rm positive\, incremental\, moduli}> 0
\end{equation}
and {\sc Hill's} inequality is satisfied if and only if $\Dabl_\lambda\widehat{\pb\tau}_1(\lambda)>0$.
This means along these special deformation states, {\sc Hill's} inequality amounts to a positive incremental modulus $\Dabl_\lambda\widehat\tau_1(\lambda)$.
Indeed, this stability concept is already used in \cite{Kos25} for the {\sc Mooney--Rivlin} model.

Along the family $\f F={\rm diag}(\lambda,\lambda^m,\lambda^{-(m+1)})$ we can always assume that the stretch in the third component adjusts itself freely due to incompressibility
once the first two components are set.
Thus, {\sc Hill's} inequality along the family $\f F={\rm diag}(\lambda,\lambda^m,\lambda^{-(m+1)})$ reads, continuing from (\ref{Hill_2D}),
\begin{equation}
0<\left<\dfrac{\partial \pb\tau}{\partial t}, \f d\right> = \underbrace{\left(\Dabl_\lambda\widehat\tau_1(\lambda)+m\,\Dabl_\lambda\widehat\tau_2(\lambda)-(m+1)\,\Dabl_\lambda\widehat\tau_3(\lambda)\right)}_{=:D^m_{1D}}\,\frac{|\dot{\lambda}|^2}{\lambda} \,.
\end{equation}
This motivates to consider the generalized incremental modulus
\begin{equation}
D^m_{1D}:=\Dabl_\lambda\widehat\tau_1(\lambda)+m\,\Dabl_\lambda\widehat\tau_2(\lambda)-(m+1)\,\Dabl_\lambda\widehat\tau_3(\lambda)\,,
\label{pos_incr_mod}
\end{equation}
and {\sc Hill's} inequality along the family $\f F={\rm diag}(\lambda,\lambda^m,\lambda^{-(m+1)})$ is satisfied, if and only if $D^m_{1D}>0$.
The stress boundary condition for $\widehat\tau_2$ and $\widehat\tau_3$ has to be incorporated, see the three prototype examples following (\ref{tau_1__tau_2}).
In our algorithmic treatment we check the latter condition for $m \in [-\frac12;1]$,
thus extending the necessary {\sc Abaqus} check from the prototype modes to the family $\f F={\rm diag}(\lambda,\lambda^m,\lambda^{-(m+1)})$.
However, the condition
\begin{equation}
D^m_{1D}>0 \quad \forall m\in[-\frac12,1]
\label{D_1D_all}
\end{equation}
does not imply {\sc Hill's} inequality for the incompressible case in general\footnote{This situation is similar to a scalar function,
which is convex along a family of straight slices which cover its domain of definition, but the function is not convex over its domain of
definition.}.
This is shown exemplary in Fig.\ \ref{InvarEbene_MoRi_Stab_2} in contrast to Fig.\ \ref{InvarEbene_MoRi_ext_HESSE_proj_2x2_c2-0p2},
where firstly (\ref{D_1D_all}) is applied above and, secondly, {\sc Hill's} inequality in the form of (\ref{HILL_final}) is used.
Nevertheless, the aim of this article is to show the application of the 1D condition (\ref{D_1D_all}) for simplified and quick stability estimations in the case of incompressible deformation modes.
%
%
%
%
\section{Another view on the {\sc Mooney--Rivlin} model (\ref{W_MoRi}): comparison of stability checks \label{app-MoRi_ext}}
This section describes a generalized version of the {\sc Mooney--Rivlin} model (\ref{W_MoRi}) in order to add statements about the convexity of this type of energy functions with $I_1$ and $I_2$.
To this end, let us consider the function
\begin{equation}
W_{\rm MoRi} = c_1\,(I_1-3)+c_2\,(I_2-3) = c_1\,(\lVert\f v\rVert^2-3)+c_2\,(\lVert\f v^{-1}\rVert^2-3)
\label{W_MoRi_ext}
\end{equation}
with the left stretch tensor $\f v=\f F\cdot \f R^{-1}$, the material parameters $c_1$, $c_2$ and the condition $\det \f v=1$ for incompressibility.
With reference to {\sc Hill}\cite{Hil70} (page 469) and reformulating the parameters $\mu$ and $\delta$ given there, one can write (\ref{W_MoRi_ext}) in the form
\begin{equation}
W_{\rm MoRi} = \frac12\mu(\frac12+\delta)\,(\lVert\f v\rVert^2-3)+\frac12\mu(\frac12-\delta)\,(\lVert\f v^{-1}\rVert^2-3)
\end{equation}
by $\mu=2(c_1+c_2)$ and $\delta=\dfrac{c_1-c_2}{2(c_1+c_2)}=\dfrac{c_1-c_2}{\mu}$ with the parameters as given in (\ref{W_MoRi_ext}).
{\sc Hill} himself (\cite{Hil70}, p.\ 496) gives the limit $|\delta| \le \frac12$ for an overall {\sc Hill's} inequality to be satisfied.
This means, the prefactors of the two energy terms, i.e.\ the material parameters, must be positive for {\sc Hill's} inequality to be satisfied.
%

\noindent
From (\ref{W_MoRi_ext}), we obtain furtheron
\begin{eqnarray}
W_{\rm MoRi} &=& c_1\,(\lVert e^{\log\f v}\rVert^2-3)+c_2\,(\lVert e^{-\log\f v}\rVert^2-3) \nonumber \\
                       &=& c_1\,([e^{2x_1}+e^{2x_2}+e^{2x_3}]-3)+c_2\,([e^{-2x_1}+e^{-2x_2}+e^{-2x_3}]-3) \nonumber \\
                      &=:& \widehat{W}(x_1,x_2,x_3) \,,
\label{W_MoRi_ext_x123}
\end{eqnarray}
where $x_i:=\log\lambda_i$, $(i=1,2,3)$ are the "principal log--stretches".
From that we directly compute the principal {\sc Kirchhoff} stresses
\begin{equation}
\tau_i = \frac{\dint\widehat{W}}{\dint x_i} = \frac{\dint\widehat{W}}{\dint\log\lambda_i}
\label{tau_log_appx}
\end{equation}
in equivalence to (\ref{tau_log}), which is used in the following development.
\subsection{Checking {\sc Hill's} inequality in the general compressible case}
\begin{figure}[!htpb]
\centering
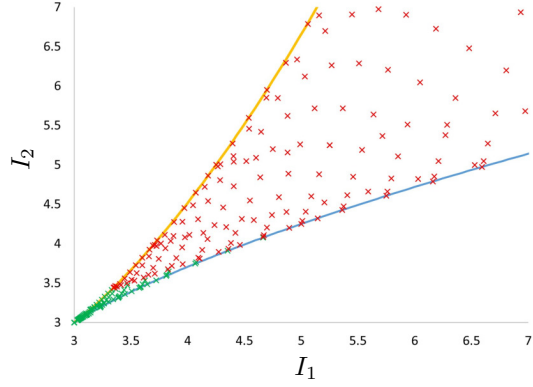 
\caption{Plane of invariants as scope for visualization the material stability of the {\sc Mooney--Rivlin} model (\ref{W_MoRi_ext_x123}) with the parameter set $c_1=0.8$~MPa and $c_2=-0.2$~MPa:
The \textcolor{green}{green} crosses here mark the stable region as defined by the condition based on positive definiteness of $\mathbb H$ in (\ref{HESSE_W_MoRi}) without taking special care of the incompressibility assumption.
\label{InvarEbene_MoRi_ext_HESSE_HILL_c2=-0.2}}
\end{figure}

Checking the convexity of $\widehat{W}$ as a function of $\log\f v$ is easy by considering the {\sc Hesse} matrix ${\mathbb H}:=\Dabl^2\widehat{W}\in{\rm Sym}(3)$.
Based on (\ref{tau_log_appx}) this step gives us the {\sc Hesse} matrix in general form
\begin{eqnarray}
{\mathbb H}&=&\begin{bmatrix} H_{11} & H_{12} & H_{13} \\ H_{12} & H_{22} & H_{23} \\ H_{13} & H_{23} & H_{33} \end{bmatrix}
= \left(\frac{\partial^2\widehat{W}(x_1,x_2,x_3)}{\partial x_i\partial x_j}\right)_{i,j=1,2,3}\\
&=&
\begin{bmatrix}
\dfrac{\partial^2\widehat{W}}{\partial x_1\partial x_1}&\dfrac{\partial^2\widehat{W}}{\partial x_1\partial x_2}&\dfrac{\partial^2\widehat{W}}{\partial x_1\partial x_3}\\[0.8em]
\dfrac{\partial^2\widehat{W}}{\partial x_2\partial x_1}&\dfrac{\partial^2\widehat{W}}{\partial x_2\partial x_2}&\dfrac{\partial^2\widehat{W}}{\partial x_2\partial x_3}\\[0.8em]
\dfrac{\partial^2\widehat{W}}{\partial x_3\partial x_1}&\dfrac{\partial^2\widehat{W}}{\partial x_3\partial x_2}&\dfrac{\partial^2\widehat{W}}{\partial x_3\partial x_3}
\end{bmatrix}
=
\begin{bmatrix}
\dfrac{\partial\tau_1}{\partial x_1}&\dfrac{\partial\tau_1}{\partial x_2}&\dfrac{\partial\tau_1}{\partial x_3}\\[0.8em]
\dfrac{\partial\tau_2}{\partial x_1}&\dfrac{\partial\tau_2}{\partial x_2}&\dfrac{\partial\tau_2}{\partial x_3}\\[0.8em]
\dfrac{\partial\tau_3}{\partial x_1}&\dfrac{\partial\tau_3}{\partial x_2}&\dfrac{\partial\tau_3}{\partial x_3}
\end{bmatrix} \quad {\rm with}\quad \pb \tau=\Dabl_{\log\f v}\widehat{W}(\log\f v)\,.  \nonumber
\label{HESSE_W_MoRi}
\end{eqnarray}
%
It is clear that if $\mathbb H$ is positive definite, the energy function $W$ is convex as a function of $\log\lambda_i$ and this convexity remains true also along the subset $\det\f F=1$ of incompressible deformations
since $1=\det\f F \Longleftrightarrow {\rm trace}(\log\f v)=0$ is a linear and convex subset in the log--space.
We may evaluate the positive definiteness of $\mathbb H$ with {\sc Sylvester's} criterion 
\begin{equation}
{H}_{11}>0, \quad {H}_{11}\cdot{H}_{22}-{H}_{12}\cdot{H}_{21} \quad {\rm and} \quad \det\mathbb H>0
\end{equation}
or
\begin{equation}
{\rm trace}(\mathbb H)>0, \quad {\rm trace}({\rm Cof}\,\mathbb H)>0\,, \quad \det\mathbb H>0
\end{equation}
along the incompressible family $\f F={\rm diag}(\lambda,\lambda^m,\lambda^{-(m+1)})$ leading with the parameter set $c_1=0.8$~MPa and $c_2=-0.2$~MPa
to Fig.\ \ref{InvarEbene_MoRi_ext_HESSE_HILL_c2=-0.2}, 
while the evaluation for $D_{1D}^m$ (from (\ref{pos_incr_mod})) gives a different result shown in Fig.\ \ref{InvarEbene_MoRi_Stab_2}.
%
%

%
%
%
%
%
\subsection{Checking {\sc Hill's} inequality in the general incompressible case: the projected {\sc Hesse} matrix $\mathbb H^{\rm inc}$}
\begin{figure}[!h]
\centering
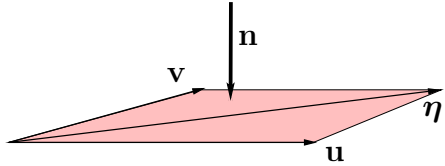 
\caption{Every vector $\pb\eta$ in the plane (\ref{plane_cond}) can be represented as $\pb\eta=\f P\cdot\pb\xi$. \label{HESSE_projection}}
\end{figure}

\begin{figure}[!h]
\centering
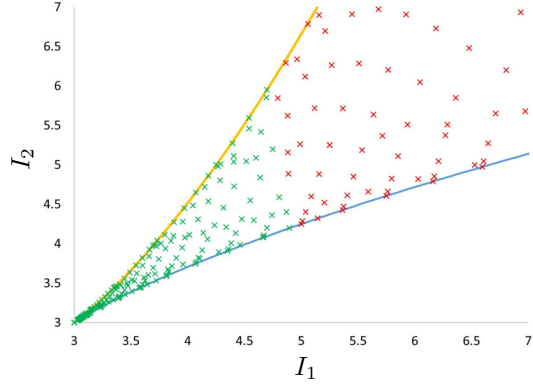 
\caption{Plane of invariants as scope for visualization the material stability of the {\sc Mooney--Rivlin} model (\ref{W_MoRi_ext_x123}) with the parameter set $c_1=0.8$~MPa and $c_2=-0.2$~MPa.
The \textcolor{green}{green} crosses here mark the complete stable region as defined by the condition based on positive definiteness of $\mathbb H^{\rm inc}$ in (\ref{HESSE_W_MoRi}),
which show a larger green (i.e.\ stable) area compared to Fig.\ \ref{InvarEbene_MoRi_ext_HESSE_HILL_c2=-0.2}  above.
\label{InvarEbene_MoRi_ext_HESSE_proj_2x2_c2-0p2}}
\end{figure}

In the log--space, the incompressibility constraint translates to $\log\lambda_1+\log\lambda_2+\log\lambda_3=0 \Leftrightarrow \lambda_1\lambda_2\lambda_3=1$.
Abbreviating $x_i=\log\lambda_i$, we need to evaluate
\begin{equation}
\left<\Dabl^2\widehat{W}(x_1,x_2,x_3)\cdot\eta, \eta\right> >0  \quad \forall\eta\neq0,\,\, \eta_1+\eta_2+\eta_3=0 \,.
\label{D2W_eta}
\end{equation}
The linear subspace
\begin{equation}
\eta_1+\eta_2+\eta_3=0
\label{plane_cond}
\end{equation}
is a plane with normal vector $\f n= \begin{bmatrix}1\\ 1\\ 1\end{bmatrix}$ as depicted in Fig.\ \ref{HESSE_projection}.
This plane is spanned by (\ref{plane_cond}) as $\f u= \begin{bmatrix}0\\ -1\\ 1\end{bmatrix}$ and $\f v= \begin{bmatrix}1\\ -1\\ 0\end{bmatrix}$.
Thus we can write any vector $\eta$ satisfying (\ref{plane_cond}) as 
\begin{equation}
\eta=\xi_1\, \f u + \xi_2\, \f v =  \begin{bmatrix}0 & 1\\ -1   & -1 \\ 1 & 0\end{bmatrix}\cdot \begin{bmatrix}\xi_1\\ \xi_2\end{bmatrix}=\f P\cdot\begin{bmatrix}\xi_1\\ \xi_2\end{bmatrix}\,.
\end{equation}
Therefore, (\ref{D2W_eta}) rewrites as 
\begin{eqnarray}
\left<\Dabl^2\widehat{W}(x_1,x_2,x_3)\cdot\eta, \eta\right>_{\mathbb R^3} &=&
\left<\Dabl^2\widehat{W}(x_1,x_2,x_3)\cdot\f P\cdot\begin{bmatrix}\xi_1\\ \xi_2\end{bmatrix}, \f P\cdot\begin{bmatrix}\xi_1\\ \xi_2\end{bmatrix}\right>_{\mathbb R^3} \\
&=& \left<\f P^{\rm T}\cdot\Dabl^2\widehat{W}(x_1,x_2,x_3)\cdot\f P\cdot\begin{bmatrix}\xi_1\\ \xi_2\end{bmatrix},\begin{bmatrix}\xi_1\\ \xi_2\end{bmatrix}\right>_{\mathbb R^2}>0  \quad \forall\,(\xi_1,\xi_2)\in{ \mathbb R^2}\ne 0 \,. \nonumber
\end{eqnarray}
Accordingly, we check the positive definiteness of the {\sl projected {\sc Hesse} matrix}
\begin{equation}
\mathbb H^{\rm inc} := \f P^{\rm T}\cdot \mathbb H\cdot\f P
=\begin{bmatrix}H_{11} - 2H_{12} + H_{22} \quad & H_{13} - H_{12} + H_{22} - H_{23} \\
                            H_{13} - H_{12} + H_{22} - H_{23} \quad & H_{22} - 2H_{23} + H_{33} \end{bmatrix}
\label{H_incompr}
\end{equation}
in the following. It is now clear, that
\begin{eqnarray}
\mathbb H^{\rm inc} \in{\rm Sym}^{++}(2) \quad &\Longleftrightarrow& \quad \left\{\mathbb H^{\rm inc}\right\}_{11}>0 \quad{\rm and}\quad\det\mathbb H^{\rm inc}>0 \nonumber \\
                                                                                &\Longleftrightarrow& \quad {\rm trace}(\mathbb H^{\rm inc})>0 \quad{\rm and}\quad\det\mathbb H^{\rm inc}>0
\label{HILL_final}
\end{eqnarray}
is {\sl the} final encompassing condition of {\sc Hill's} inequality in the incompressible case, see Fig.~\ref{InvarEbene_MoRi_ext_HESSE_proj_2x2_c2-0p2}.
Subsequently we will show that this condition coincides with the {\sc Abaqus} check $\mathbb D\in {\rm Sym}^{++}(2)$ of (\ref{D_mat}).
Note again that $\mathbb H^{\rm inc} \in{\rm Sym}^{++}(2)$ is sufficient for $D_{1D}^m>0$, but not the other way round, implying that the stability limits calculated
based on  $D_{1D}^m>0$ are larger than those calculated for $\mathbb H^{\rm inc} \in{\rm Sym}^{++}(2)$, see (\ref{stab_relations}) and
Table \ref{tab_stable_eps}\footnote{As will be seen, {\sc Hill's} incompressible inequality condition $\mathbb H^{\rm inc} \in{\rm Sym}^{++}(2)$ (*) is equivalent to the
{\sc Abaqus Drucker} condition.
This can be evaluated for the {\sc Mooney--Rivlin} model in certain deformation states, e.g.\ a deformation shown in uniaxial tension.
There, no use is made of stress boundary conditions, i.e.\ it is irrelevant, how we got uniaxial tension geometry.
On the other hand, another condition is e.g.\ looking at uniaxial tension, incorporating the stress condition, and then requiring monotonicity of {\sc Cauchy} stress versus stretch: $D^m_{1D}>0$ (**).
Now it holds  (*) $\Rightarrow$ (**), but not the other way round.
Therefore, the limits calculated from (**) are larger than the limits calculated from (*), while both have been calculated for the geometry of uniaxial tension, see Table \ref{tab_stable_eps}.}.
%
%
%
%
\clearpage

\begin{figure}[!h]
\centering
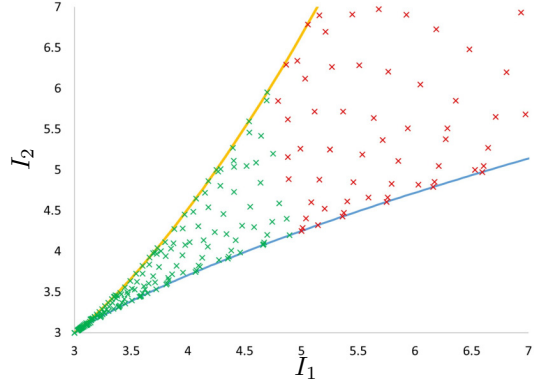 
\caption{Plane of invariants as scope for visualization the material stability of the {\sc Mooney--Rivlin} model (\ref{W_MoRi_ext_x123}) with the parameter set $c_1=0.8$~MPa and $c_2=-0.2$~MPa.
The \textcolor{green}{green} crosses here mark the complete stable region as defined by the condition based on positive definiteness of $\mathbb D$ in (\ref{D_mat})
showing exactly the same result as Fig.\ \ref{InvarEbene_MoRi_ext_HESSE_proj_2x2_c2-0p2}  above, based on $\mathbb H^{\rm inc}$.
\label{InvarEbene_MoRi_D_ABAQ}}
\end{figure}

\begin{figure}[!h]
\centering
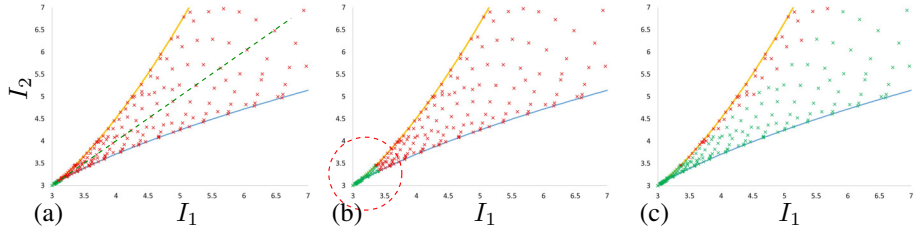
\caption{Summary for the {\sc Mooney--Rivlin} model with the parameter set $c_1=1$~MPa and $c_2=$\textcolor{red}{-0.4} MPa
based on (a) $\mathbb H$ in (\ref{HESSE_W_MoRi}), based on (b) $\mathbb H^{\rm inc}$ in (\ref{H_incompr}) and on (c) $D_{1D}^m$ in (\ref{D_1D_m}).
\label{InvarEbene_MoRi_Stab__summary_c2=-0p4}}
\end{figure}

In contrast, the built--in {\sc Abaqus} stability check gives a stable situation for the three prototype deformations and the {\sc Mooney--Rivlin} model using the discussed
parameter set $c_1=0.8$~MPa and $c_2=-0.2$~MPa in the range of
\begin{eqnarray}
\varepsilon_{\rm uniax}&\in&(-0.5614 ... 1.00) \Longleftrightarrow \lambda\in(0.4386 ... 2)    \,, \nonumber \\
\varepsilon_{\rm pure\, shear}&\in&(-0.4737 ... 0.99) \Longleftrightarrow \lambda\in(0.5263 ... 1.9) \,, \nonumber \\
\varepsilon_{\rm biax}&\in&(-0.2919 ... 0.51)  \Longleftrightarrow \lambda\in(0.7071 ... 1.51) \nonumber
\end{eqnarray}
for the nominal strains $\varepsilon=\lambda-1$. The output of {\sc Abaqus} material (stability) evaluation is given by
\begin{verbatim}
     HYPERELASTICITY - MOONEY-RIVLIN STRAIN ENERGY

            C1       C2
            0.8      -0.2    

 ***WARNING: UNSTABLE HYPERELASTIC MATERIAL

    UNIAXIAL TENSION:        UNSTABLE AT A NOMINAL STRAIN LARGER THAN        1.0000
    UNIAXIAL COMPRESSION:    UNSTABLE AT A NOMINAL STRAIN LESS THAN         -0.5614
    BIAXIAL TENSION:         UNSTABLE AT A NOMINAL STRAIN LARGER THAN        0.5100
    BIAXIAL COMPRESSION:     UNSTABLE AT A NOMINAL STRAIN LESS THAN         -0.2929
    PLANAR TENSION:          UNSTABLE AT A NOMINAL STRAIN LARGER THAN        0.9000
    PLANAR COMPRESSION:      UNSTABLE AT A NOMINAL STRAIN LESS THAN         -0.4737
\end{verbatim}
whereas the here given boundary values $\varepsilon_{\rm uniax}=-0.5614$, $\varepsilon_{\rm pure\, shear}=\varepsilon_{\rm planar\, compression}=-0.4737$ and
$\varepsilon_{\rm biax}=-0.2929$ can be exactly reproduced by (\ref{D_mat}) for the {\sc Mooney--Rivlin} model (\ref{W_MoRi})
applying the coefficients of $\mathbb D$ to obtain $\lambda=\varepsilon_\star+1$ for that model in the form
\begin{eqnarray}
D_{11} &=&  4(\lambda_1^2+\lambda_3^2)(c_1+\lambda_2^2c_2)\,,         \nonumber \\
D_{22} &=&  4(\lambda_2^2+\lambda_3^2)(c_1+\lambda_1^2c_2)\,,          \nonumber \\
D_{12} = D_{21} &=& 4\lambda_3^2c_1+4\lambda_3^{-2}c_2
\end{eqnarray}
directly as given in \cite{Aba09} for the incompressible case for any strain energy density functions $W=W(\bar{I}_1,\bar{I}_2)$
\begin{eqnarray}
D_{11} &=& 4(\lambda_1^2+\lambda_3^2)(\frac{\partial W}{\partial\bar{I_1}}+\lambda_2^2\frac{\partial W}{\partial\bar{I_2}})
+ 4(\lambda_1^2-\lambda_3^2)^2(\frac{\partial^2 W}{\partial\bar{I_1}^2}+2\lambda_2^2\frac{\partial^2 W}{\partial\bar{I_1}\partial\bar{I_2}}+\lambda_2^4\frac{\partial^2 W}{\partial\bar{I_2}^2})\,, \nonumber \\
D_{22} &=& 4(\lambda_2^2+\lambda_3^2)(\frac{\partial W}{\partial\bar{I_1}}+\lambda_1^2\frac{\partial W}{\partial\bar{I_2}})
+ 4(\lambda_2^2-\lambda_3^2)^2(\frac{\partial^2 W}{\partial\bar{I_1}^2}+2\lambda_1^2\frac{\partial^2 W}{\partial\bar{I_1}\partial\bar{I_2}}+\lambda_1^4\frac{\partial^2 W}{\partial\bar{I_2}^2})\,, \label{app_D_altern} \\
D_{12} = D_{21} &=& 4\lambda_3^2\frac{\partial W}{\partial\bar{I_1}}+4\lambda_3^{-2}\frac{\partial W}{\partial\bar{I_2}}
+ 4(\lambda_1^2-\lambda_3^2)(\lambda_2^2-\lambda_3^2)(\frac{\partial^2 W}{\partial\bar{I_1}^2}+(\lambda_1^2+\lambda_2^2)\frac{\partial^2 W}{\partial\bar{I_1}\partial\bar{I_2}}+\lambda_1^2\lambda_2^2\frac{\partial^2 W}{\partial\bar{I_2}^2})  \nonumber\,.
\end{eqnarray}
Checking $\mathbb D\in{\rm Sym^{++}(2)}$ based on (\ref{app_D_altern}) results in Fig.\ \ref{InvarEbene_MoRi_D_ABAQ}.
Summarizing, we give an overview in the following Tab.\ \ref{tab_stable_eps}.
%
%
\begin{table}[!h]
\begin{tabular}{lllll}
                                               &uniax mode & pure shear & biax	\\[2mm]
%
(\ref{HESSE_W_MoRi}): $\mathbb H$ \quad $\in {\rm Sym}^{++}(3)$ & $\lambda\in(0.708 ... 1.999)$ & $\lambda\in(0.7072 ... 1.4142)$ &$\lambda\in(0.708 ... 1.189)$ \\[1 mm]
%
(\ref{H_incompr}): $\mathbb H^{\rm inc}$\! $\in {\rm Sym}^{++}(2)$ & $\lambda\in(0.4445 ... 1.999)$ & $\lambda\in(0.528 ... 1.893)$ & $\lambda\in(0.708 ... 1.5)$ \\[1 mm]
%
\,\,\,(\ref{D_mat}): $\mathbb D$ \quad $\in {\rm Sym}^{++}(2)$ & $\lambda\in(0.445 ... 1.999)$ & $\lambda\in(0.53 ... 1.894)$ & $\lambda\in(0.7072 ... 1.5005)$ \\[1 mm]
%
(\ref{pos_incr_mod}): $D_{1D}^m$ $>0$ & $\lambda\in(0.4442 ... \infty)$ & $\lambda\in(0 ... \infty)$ & $\lambda\in(0 ... 1.5)$ \\[1 mm]
%
%
%
%
{\sc Abaqus} built--in &  $\lambda\in(0.4386 ... 2)$ & $\lambda\in(0.5263 ... 1.9)$ & $\lambda\in(0.7071 ... 1.51)$ \\[3mm]
%
%
%
\end{tabular}
\caption{Comparison of the different stability ranges of the {\sc Mooney--Rivlin} model with $c_1=0.8$ MPa and $c_2=-0.2$ MPa for the various criteria from above wrt.\ the stretch $\lambda=\varepsilon+1$ and the three prototype modes $m=-\frac12, 0 , 1$.
Please note that regarding the accuracies of the third and fourth decimal places, it is evident that {\sc Abaqus} also uses a significantly higher accuracy
than $\Delta \lambda=0.01$ for this evaluation than is specified in the manual and previously cited in Sec.\ \ref{sec:Background}.
While the lower and the upper stability limits based on $\mathbb H^{\rm inc}$, $\mathbb D$ and {\sc Abaqus} must be identical, different numerical implementations lead --  in the view of the author -- to slightly different values.
\label{tab_stable_eps}}
\end{table}
With respect to Tab.\ \ref{tab_stable_eps}, and ignoring numerical errors, we observe for the three prototype modes $m=-\frac12, 0, 1$ the relations
\begin{align}
\label{stab_relations}
\mathbb H&\in {\rm Sym}^{++}(3) \\
&\Downarrow \nonumber \\
\mathbb H^{\rm inc} &\in {\rm Sym}^{++}(2) \Longrightarrow {\mathbb D} \in {\rm Sym}^{++}(2) &\nonumber \\
&\hspace*{26mm} \Updownarrow \nonumber \\
&\hspace*{24mm} {\textrm {\sc Abaqus}\, {\rm built\!\!-\!\!in}\, {\rm check}}  \Longrightarrow D_{1D}^m>0 \,. \nonumber
\end{align}
To show a counter example for the {\sc Mooney--Rivlin} model, we give an evaluation of the three stability criteria for that model with the parameter set
$c_1=0.4$ MPa and $c_2=$\textcolor{red}{-0.4} MPa in Fig.\ \ref{InvarEbene_MoRi_Stab__summary_c2=-0p4}, where for the three criteria each gives a different picture. 

In addition to the discussion on the {\sc Mooney--Rivlin} model above, here we summarize the representations of the {\sc Yeoh} model from Section \ref{sec:Matlab_Toolbox} and
supplement them with the derivations here concerning $\mathbb H$ of (\ref{HESSE_W_MoRi}) and $\mathbb H^{\rm inc}$ of (\ref{H_incompr}).
\begin{figure}[!h]
\centering
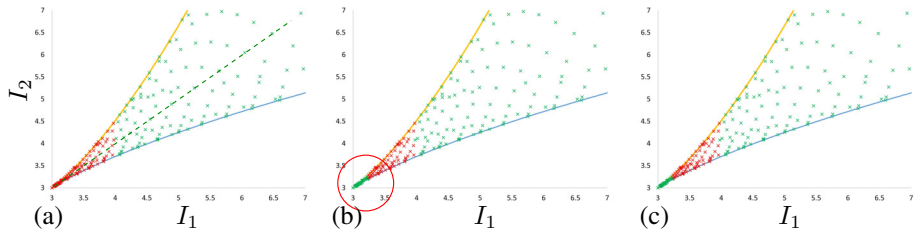
\caption{Summary for the {\rm Yeoh} model with the parameter set $c_1=1$~MPa, $c_2=-0.9$~MPa and $c_3=0.3$~MPa
based on (a) $\mathbb H$ in (\ref{HESSE_W_MoRi}), based on (b) $\mathbb H^{\rm inc}$ in (\ref{H_incompr}) 
and on (c) $D_{1D}^m$ in (\ref{D_1D_m}), where (b) is equivalent to Fig.\ \ref{InvarEbene_Yeoh_Stab__ABAQ_ganzeEbene} and (c) is equivalent to Fig.\ \ref{InvarEbene_Yeoh_Stab}.
\label{InvarEbene_Yeoh_Stab__summary}}
\end{figure}
%
%
%
%
%
\subsection{Equivalence of positive definiteness for $\mathbb H^{\rm inc}$ and $\mathbb D$ \label{app_HH_DD}}
As seen in the previous section,
for the three prototype modes, the positive definiteness of $\mathbb H^{\rm inc}$ and of $\mathbb D$ give the same stability limits.
However, for a general strain state, the relation between $\mathbb H^{\rm inc}$ and $\mathbb D$ is not clear at all at first sight.

On the one hand, we have from (\ref{H_incompr}) 
\begin{eqnarray}
\mathbb H^{\rm inc} &:=& \f P^{\rm T}\cdot \mathbb H\cdot\f P = \f P^{\rm T}\cdot \Dabl^2_{\log\lambda_i}\widehat{W} \cdot\f P \\
&=&\begin{bmatrix}H_{22} - 2H_{23} + H_{33} \quad & H_{13} - H_{12} + H_{22} - H_{23} \\
                            H_{13} - H_{12} + H_{22} - H_{23} \quad & H_{11} - 2H_{12} + H_{22} \end{bmatrix}
\quad {\rm with} \quad\f P=\begin{bmatrix}0 & 1\\ -1   & -1 \\ 1 & 0\end{bmatrix}\,. \nonumber
\label{Hessian_reduced}
\end{eqnarray}

\noindent
On the other hand, it holds by assuming $\tau_2=0$ that (surprisingly, formula from {\sc Patrizio Neff}, {\sl private communication})
\begin{eqnarray}
\widehat{\tau}_1 &=& \partial_{\log\lambda_1} [W_{\rm red}^{\rm inc}(\lambda_1,\lambda_3)]\,, \nonumber \\
\widehat{\tau}_3 &=& \partial_{\log\lambda_3} [W_{\rm red}^{\rm inc}(\lambda_1,\lambda_3)]
\end{eqnarray}
so that
\begin{eqnarray}
\overline{\mathbb D}&=&\begin{bmatrix}\Dabl_{\log \lambda_1}\, \widehat{\tau}_1(\log \lambda_1,\log \lambda_3) \quad \Dabl_{\log \lambda_3}\, \widehat{\tau}_1(\log \lambda_1,\log \lambda_3) \\ 
                                                   \Dabl_{\log \lambda_1}\, \widehat{\tau}_3(\log \lambda_1,\log \lambda_3) \quad \Dabl_{\log \lambda_3}\, \widehat{\tau}_3(\log \lambda_1,\log \lambda_3) \end{bmatrix}\nonumber\\
&=&  \Dabl^2_{\log\lambda_i}[W_{\rm red}^{\rm inc}(\lambda_1,\lambda_3)] = \Dabl^2_{\log\lambda_i}[\widehat{W}_{\rm red}^{\rm inc}(\log\lambda_1,\log\lambda_3)]\,,
\end{eqnarray}
where we have to set
\begin{eqnarray}
W_{\rm red}^{\rm inc}(\lambda_1,\lambda_3)&:=&W(\lambda_1,\frac{1}{\lambda_1\lambda_3},\lambda_3)\,, \nonumber\\
\widehat{W}_{\rm red}^{\rm inc}(\log\lambda_1,\log\lambda_3)&:=&W_{\rm red}^{\rm inc}(e^{\log\lambda_1}, e^{\log\lambda_3}) \nonumber\\
&=&W(e^{\log\lambda_1}, e^{-(\log\lambda_1+\log\lambda_3)}, e^{\log\lambda_3})
\label{def_W__inc_red}
\end{eqnarray}
because of $\lambda_1^{-1}\lambda_3^{-1}=e^{-\log\lambda_1}\, e^{-\log\lambda_3}=e^{-(\log\lambda_1+\log\lambda_3)}$.
Let us write 
\begin{equation}
\widehat{W}_{\rm red}^{\rm inc}(x_1,x_3):=W(e^{x_1}, e^{-(x_1+x_3)}, e^{x_3}) = \widehat{W}(x_1,-(x_1+x_3),x_3)\,,
\end{equation}
which gives for the {\sc Hesse} matrix of $\widehat{W}^{\rm inc}_{\rm red}$
\begin{equation}
\Dabl^2_{(x_1,x_3)}\widehat{W}_{\rm red}^{\rm inc}(x_1,x_3)=
\begin{bmatrix}
\delta_1\delta_1\widehat{W} - 2\delta_1\delta_2\widehat{W} + \delta_2\delta_2\widehat{W} & \delta_1\delta_3\widehat{W} - \delta_1\delta_2\widehat{W} - \delta_2\delta_3\widehat{W} + \delta_2\delta_2\widehat{W} \\
\delta_1\delta_3\widehat{W} - \delta_1\delta_2\widehat{W} - \delta_2\delta_3\widehat{W} + \delta_2\delta_2\widehat{W} & \delta_2\delta_2\widehat{W} - 2\delta_2\delta_3\widehat{W} + \delta_3\delta_3\widehat{W}
\end{bmatrix} \,,
\label{Hessian_direct}
\end{equation}
so that 
\begin{equation}
\label{D_quer_mat}
\overline{\mathbb D} = \begin{bmatrix}H_{11} - 2H_{12} + H_{22} \quad & H_{13} - H_{12} - H_{23} + H_{22} \\
                                                               H_{13} - H_{12} - H_{23} + H_{22} \quad & H_{22} - 2H_{23} + H_{33} \end{bmatrix}\,.
\end{equation}
%
%
Now, for isotropic response, the elastic energy $W$ is invariant under permutations of the principal stretches. Therefore, the positive definiteness of $\mathbb D$ from (\ref{D_mat})
and the positive definiteness of $\overline{\mathbb D}$ from (\ref{D_quer_mat}) are equivalent.
Indeed, on the level of $\mathbb D$ in (\ref{D_mat}) it is an arbitrary choice to require $\tau_3=0$. Here, we use instead $\tau_2=0$ for consistency.
Comparing (\ref{D_quer_mat}) and (\ref{Hessian_reduced}) we have shown that
\begin{equation}
\mathbb H^{\rm inc}\in {\rm Sym}^{++}(2) \Longleftrightarrow \overline{\mathbb D}\in {\rm Sym}^{++}(2)
\end{equation}
since ${\rm trace}(\mathbb H^{\rm inc})={\rm trace}(\overline{\mathbb D})$ and $\det\mathbb H^{\rm inc}= \det\overline{\mathbb D}$.
Thus
\begin{equation}
\mathbb H^{\rm inc}\in {\rm Sym}^{++}(2) \quad \Longleftrightarrow \quad \mathbb D\in {\rm Sym}^{++}(2)
\end{equation}
and the {\sc Abaqus} stability check is shown to be fully equivalent to {\sc Hill's} condition for the incompressible case.
%
%
%
%
%
\section{Non--monotonicity of the log shear measure in {\sl simple shear} \label{app-B_SiSh}}
%
%
Already for {\sl simple shear} it can be shown that the shear components of the logarithmic strain tensor themselves exhibit a non--monotonic behavior with respect to the deformation $\gamma=\tan\alpha$,
see Fig.\ \ref{SiSh_logb12} below. Here we give a closed form and derivation of $(\log \f b)_{12}$ in {\sl simple shear} as function of $\gamma$.

{Starting with the unique {\sl polar decomposition} of the deformation gradient
\begin{equation}
\f F = \f v\cdot \f R = \begin{bmatrix}1 \quad \gamma \quad 0 \\ 0 \quad 1 \quad 0\\ 0 \quad 0 \quad 1\end{bmatrix}
\label{F_vR}
\end{equation}
in the {\sl simple shear} case by an eigenvalue decomposition of
\begin{equation}
\f b = \f F\cdot \f F^{\rm T} = \begin{bmatrix}1+\gamma^2 \quad &\gamma& \quad 0 \\ \gamma \quad &1& \quad 0\\ 0  \quad &0& \quad 1\end{bmatrix}
\label{b_tensor}
\end{equation}
by $(\f b - \mu\f I)\cdot \f n=\f 0$ gives the eigenvalues of $\f b$ as
\begin{equation}
\mu_{1,2} = \left(\frac{\gamma \pm \sqrt{\gamma^2+4}}{2}\right)^2, \quad \mu_3=1\,,
\end{equation}
see \cite{Its09}, and the three principal spatial directions
\begin{equation}
\f n_1=
\begin{bmatrix}
\dfrac{2}{\sqrt{2\gamma^2-2\gamma\sqrt{\gamma^2+4\,\,}+8\,\,}}\\ \dfrac{-\gamma+\sqrt{\gamma^2+4\,\,}}{\sqrt{2\gamma^2-2\gamma\sqrt{\gamma^2+4\,\,}+8\,\,}}\\ 0
\end{bmatrix}, \quad
\f n_2=
\begin{bmatrix}
\dfrac{2}{\sqrt{2\gamma^2+2\gamma\sqrt{\gamma^2+4\,\,}+8\,\,}}\\ \dfrac{-\gamma-\sqrt{\gamma^2+4\,\,}}{\sqrt{2\gamma^2+2\gamma\sqrt{\gamma^2+4\,\,}+8\,\,}}\\ 0
\end{bmatrix} \quad {\rm and} \quad \f n_3=\begin{bmatrix}0 \\ 0\\ 1\end{bmatrix},
\label{n_von_b}
\end{equation}
which are mutually orthogonal and normalized, see \cite{Hol00} and \cite{Its09}.
From the author's point of view, this compact representation (\ref{n_von_b}) is also rarely found in literature, while the decomposition for $\f C=\f F^{\rm T}\cdot\f F$ in contrast to $\f b$
(\ref{b_tensor}) is often seen.}

{This results in the {\sl spectral decomposition}
\begin{equation}
\f b =\f v^2=\mu_1\, \f n_1\otimes\f n_1+\mu_2\, \f n_2\otimes\f n_2+\mu_3\, \f n_3\otimes\f n_3\,,
\label{b_v2_specdecomp}
\end{equation}
which allows us to compute the matrix logarithm operation on $\f b$ by
\begin{equation}
\log\f b =\log(\mu_1)\,\f n_1\otimes\f n_1+\log(\mu_2)\,\f n_2\otimes\f n_2+\log(\mu_3)\,\f n_3\otimes\f n_3\,.
\label{log_b_specdecomp}
\end{equation}
\begin{figure}[!ht]
\centering
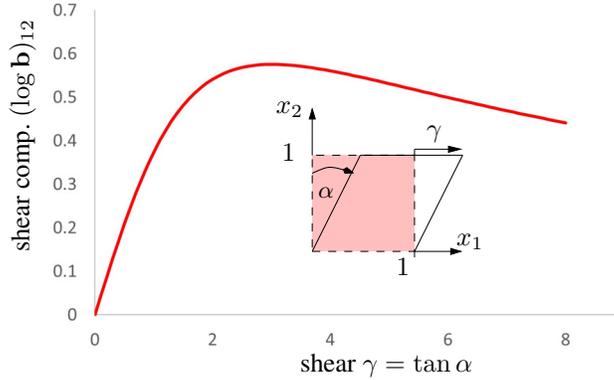   
\caption{Shear component $(\log \f b)_{12}$ as function of shear measure $F_{12}=\gamma=\tan\alpha$, which in this context is by no means a monotonous function.}
\label{SiSh_logb12}
\end{figure}
The analytical derivation using {\sc Matlab Symbolic Toolbox} \cite{MATLAB24} leads directly to the closed form expression
\begin{equation}
(\log\f b)_{12}=\frac{\log(\gamma\sqrt{\gamma^2+4}+\frac{\gamma^3}{2}\sqrt{\gamma^2+4}+2\gamma^2+\frac{\gamma^4}{2}+1)}{\sqrt{\gamma^2+4}}
\label{log_b_12}
\end{equation}
for the 12--shear component of (\ref{log_b_specdecomp}) as function of the shear measure $\gamma$ in (\ref{F_vR}).
%
%
The graphical representation of (\ref{log_b_12}) is given in Fig.\ \ref{SiSh_logb12} and shows a drop in the curve after $\gamma\simeq 3$,
which indicates a non--monotonicity even in the $\log$ strain as function of the shear measure $\gamma$.
Thus, {\sc Hencky's} quadratic model $\pb \sigma=\mu\,\log\f b$ satisfies {\sc Hill's} inequality, but predicts a non--monotone shear--stress curve $\tau_{12}=\mu\,(\log\f b)_{12}$.
}
\end{appendix}
%
%
%
\bibliographystyle{spmpsci}      
%
\bibliography{jourfull,HBlit}

%
%

\end{document}

%% file: Abaq_Check_InvarEbene.pdf_t
\begin{picture}(0,0)%
\includegraphics{Abaq_Check_InvarEbene.pdf}%
\end{picture}%
\setlength{\unitlength}{3315sp}%
\begin{picture}(5736,3108)(751,-3314)
\put(991,-1726){\rotatebox{90.0}{\makebox(0,0)[lb]{\smash{\fontsize{10}{12}\usefont{T1}{ptm}{m}{n}{\color[rgb]{0,0,0}$I_2$}%
}}}}
\put(2971,-3256){\makebox(0,0)[lb]{\smash{\fontsize{10}{12}\usefont{T1}{ptm}{m}{n}{\color[rgb]{0,0,0}$I_1$}%
}}}
\put(4051,-1096){\makebox(0,0)[lb]{\smash{\fontsize{10}{12}\usefont{T1}{ptm}{m}{n}{\color[rgb]{0,.56,0}$m=0$}%
}}}
\put(766,-3121){\makebox(0,0)[lb]{\smash{\fontsize{8}{9.6}\usefont{T1}{ptm}{m}{n}{\color[rgb]{0,0,0}$\lambda=1$}%
}}}
\put(2296,-2626){\makebox(0,0)[lb]{\smash{\fontsize{8}{9.6}\usefont{T1}{ptm}{m}{n}{\color[rgb]{0,0,1}$\lambda=2$}%
}}}
\put(4051,-1276){\makebox(0,0)[lb]{\smash{\fontsize{10}{12}\usefont{T1}{ptm}{m}{n}{\color[rgb]{0,.56,0}all shear deformations with $I_1=I_2$}%
}}}
\put(4051,-1636){\makebox(0,0)[lb]{\smash{\fontsize{9}{10.8}\usefont{T1}{ptm}{m}{n}{\color[rgb]{0,.56,0}or e.g.\ {\sl simple shear}, see (\ref{F_vR})$_2$}%
}}}
\put(3826,-2176){\makebox(0,0)[lb]{\smash{\fontsize{9}{10.8}\usefont{T1}{ptm}{m}{n}{\color[rgb]{0,0,1}uniaxial deformation: $m=-\frac12$}%
}}}
\put(3826,-2356){\makebox(0,0)[lb]{\smash{\fontsize{8}{9.6}\usefont{T1}{ptm}{m}{n}{\color[rgb]{0,0,1}$\f F={\rm diag}(\lambda, 1/\sqrt{\lambda}, 1/\sqrt{\lambda})$}%
}}}
\put(4051,-1456){\makebox(0,0)[lb]{\smash{\fontsize{9}{10.8}\usefont{T1}{ptm}{m}{n}{\color[rgb]{0,.56,0}{\sl pure shear} with $\f F={\rm diag}(\lambda, 1, \lambda^{-1})$}%
}}}
\put(1621,-511){\makebox(0,0)[lb]{\smash{\fontsize{9}{10.8}\usefont{T1}{ptm}{m}{n}{\color[rgb]{1,0,0}$\f F={\rm diag}(\lambda, \lambda, \lambda^{-2})$}%
}}}
\put(1621,-331){\makebox(0,0)[lb]{\smash{\fontsize{10}{12}\usefont{T1}{ptm}{m}{n}{\color[rgb]{1,0,0}biaxial deformation: $m=1$}%
}}}
\put(2491,-2121){\makebox(0,0)[lb]{\smash{\fontsize{8}{9.6}\usefont{T1}{ptm}{m}{n}{\color[rgb]{0,.56,0}$\lambda=2$}%
}}}
\put(2851,-936){\makebox(0,0)[lb]{\smash{\fontsize{8}{9.6}\usefont{T1}{ptm}{m}{n}{\color[rgb]{0,0,0}$\lambda=2$}%
}}}
\put(2851,-1081){\makebox(0,0)[lb]{\smash{\fontsize{8}{9.6}\usefont{T1}{ptm}{m}{n}{\color[rgb]{0,0,0}iso line}%
}}}
\put(1585,-1638){\makebox(0,0)[lb]{\smash{\fontsize{8}{9.6}\usefont{T1}{ptm}{m}{n}{\color[rgb]{1,0,0}$\lambda=1.5$}%
}}}
\end{picture}%

%% file: biax_uniax_equivalence.pdf_t
\begin{picture}(0,0)%
\includegraphics{biax_uniax_equivalence.pdf}%
\end{picture}%
\setlength{\unitlength}{4144sp}%
\begin{picture}(2794,1928)(1082,-1905)
\put(1208,-1851){\makebox(0,0)[lb]{\smash{\fontsize{9}{10.8}\usefont{T1}{ptm}{m}{n}{\color[rgb]{0,0,0}$x_3$}%
}}}
\put(2706,-118){\makebox(0,0)[lb]{\smash{\fontsize{9}{10.8}\usefont{T1}{ptm}{m}{n}{\color[rgb]{0,0,0}$x_2$}%
}}}
\put(3781,-1256){\makebox(0,0)[lb]{\smash{\fontsize{9}{10.8}\usefont{T1}{ptm}{m}{n}{\color[rgb]{0,0,0}$x_1$}%
}}}
\put(3011,-1497){\makebox(0,0)[lb]{\smash{\fontsize{9}{10.8}\usefont{T1}{ptm}{m}{n}{\color[rgb]{0,0,0}$\lambda_{\rm uniax}$}%
}}}
\put(3441,-360){\makebox(0,0)[lb]{\smash{\fontsize{9}{10.8}\usefont{T1}{ptm}{m}{n}{\color[rgb]{0,0,0}$\lambda_{\rm biax}$}%
}}}
\end{picture}%

%% file: m_equiv.pdf_t
\begin{picture}(0,0)%
\includegraphics{m_equiv.pdf}%
\end{picture}%
\setlength{\unitlength}{4144sp}%
\begin{picture}(2811,1620)(1194,-1771)
\put(3691,-1096){\makebox(0,0)[lb]{\smash{\fontsize{10}{12}\usefont{T1}{ptm}{m}{n}{\color[rgb]{0,0,0}$m$}%
}}}
\put(2431,-331){\makebox(0,0)[lb]{\smash{\fontsize{10}{12}\usefont{T1}{ptm}{m}{n}{\color[rgb]{0,0,0}$m^\star$}%
}}}
\put(2412,-1077){\makebox(0,0)[lb]{\smash{\fontsize{9}{10.8}\usefont{T1}{ptm}{m}{n}{\color[rgb]{0,1,0}\sl pure shear}%
}}}
\put(3601,-1448){\makebox(0,0)[lb]{\smash{\fontsize{9}{10.8}\usefont{T1}{ptm}{m}{n}{\color[rgb]{0,1,0}\sl biax}%
}}}
\put(1209,-646){\makebox(0,0)[lb]{\smash{\fontsize{9}{10.8}\usefont{T1}{ptm}{m}{n}{\color[rgb]{0,1,0}\sl uniaxial}%
}}}
\end{picture}%

%% file: biax-tester.pdf_t
\begin{picture}(0,0)%
\includegraphics{biax-tester.pdf}%
\end{picture}%
\setlength{\unitlength}{4144sp}%
\begin{picture}(3105,2430)(361,-2806)
\put(2746,-1186){\makebox(0,0)[lb]{\smash{\fontsize{11}{13.2}\usefont{T1}{ptm}{m}{n}{\color[rgb]{1,1,1}$x_1$}%
}}}
\put(1756,-781){\makebox(0,0)[lb]{\smash{\fontsize{11}{13.2}\usefont{T1}{ptm}{m}{n}{\color[rgb]{1,1,1}$x_2$}%
}}}
\end{picture}%

%% file: example_monotonicity.pdf_t
\begin{picture}(0,0)%
\includegraphics{example_monotonicity.pdf}%
\end{picture}%
\setlength{\unitlength}{4144sp}%
\begin{picture}(3596,2037)(1306,-2188)
\put(1494,-1824){\rotatebox{90.0}{\makebox(0,0)[lb]{\smash{\fontsize{10}{12}\usefont{T1}{ptm}{m}{n}{\color[rgb]{0,0,0}uniaxial true stress $\sigma$}%
}}}}
\put(1936,-2131){\makebox(0,0)[lb]{\smash{\fontsize{9}{10.8}\usefont{T1}{ptm}{m}{n}{\color[rgb]{0,0,0}uniaxial true strain $\epsilon^{\rm log}=\log \lambda$}%
}}}
\put(3286,-1231){\makebox(0,0)[lb]{\smash{\fontsize{9}{10.8}\usefont{T1}{ptm}{m}{n}{\color[rgb]{0,.69,0}$D_{\rm 1D}>0$}%
}}}
\put(3196,-1546){\makebox(0,0)[lb]{\smash{\fontsize{9}{10.8}\usefont{T1}{ptm}{m}{n}{\color[rgb]{1,0,0}$D_{\rm 1D}<0$}%
}}}
\put(4321,-781){\makebox(0,0)[lb]{\smash{\fontsize{10}{12}\usefont{T1}{ptm}{m}{n}{\color[rgb]{1,0,0}unstable}%
}}}
\put(4321,-961){\makebox(0,0)[lb]{\smash{\fontsize{10}{12}\usefont{T1}{ptm}{m}{n}{\color[rgb]{1,0,0}{\sc Yeoh} example}%
}}}
\put(2251,-736){\makebox(0,0)[lb]{\smash{\fontsize{10}{12}\usefont{T1}{ptm}{m}{n}{\color[rgb]{0,.69,0}{\sc Yeoh} example}%
}}}
\put(2251,-556){\makebox(0,0)[lb]{\smash{\fontsize{10}{12}\usefont{T1}{ptm}{m}{n}{\color[rgb]{0,.69,0}stable}%
}}}
\end{picture}%

%% file: InvarEbene_Yeoh_Stab.pdf_t
\begin{picture}(0,0)%
\includegraphics{InvarEbene_Yeoh_Stab.pdf}%
\end{picture}%
\setlength{\unitlength}{3729sp}%
\begin{picture}(5699,3070)(860,-3314)
\put(991,-1726){\rotatebox{90.0}{\makebox(0,0)[lb]{\smash{\fontsize{10}{12}\usefont{T1}{ptm}{m}{n}{\color[rgb]{0,0,0}$I_2$}%
}}}}
\put(2971,-3256){\makebox(0,0)[lb]{\smash{\fontsize{10}{12}\usefont{T1}{ptm}{m}{n}{\color[rgb]{0,0,0}$I_1$}%
}}}
\put(2701,-2491){\makebox(0,0)[lb]{\smash{\fontsize{10}{12}\usefont{T1}{ptm}{m}{n}{\color[rgb]{0,0,1}limitation uniaxial deformation: $m=-\frac12$}%
}}}
\put(2701,-2671){\makebox(0,0)[lb]{\smash{\fontsize{9}{10.8}\usefont{T1}{ptm}{m}{n}{\color[rgb]{0,0,1}$\f F={\rm diag}(\lambda, 1/\sqrt{\lambda}, 1/\sqrt{\lambda})$}%
}}}
\put(1486,-376){\makebox(0,0)[lb]{\smash{\fontsize{10}{12}\usefont{T1}{ptm}{m}{n}{\color[rgb]{0,0,1}biaxial deformation}%
}}}
\put(1486,-556){\makebox(0,0)[lb]{\smash{\fontsize{10}{12}\usefont{T1}{ptm}{m}{n}{\color[rgb]{0,0,1}$m=1$}%
}}}
\put(1486,-736){\makebox(0,0)[lb]{\smash{\fontsize{9}{10.8}\usefont{T1}{ptm}{m}{n}{\color[rgb]{0,0,1}$\f F={\rm diag}(\lambda, \lambda, \lambda^{-2})$}%
}}}
\put(4006,-1276){\makebox(0,0)[lb]{\smash{\fontsize{10}{12}\usefont{T1}{ptm}{m}{n}{\color[rgb]{0,.56,0}{\sl pure shear} with $\f F={\rm diag}(\lambda,1,\lambda^{-1})$}%
}}}
\put(4006,-1096){\makebox(0,0)[lb]{\smash{\fontsize{10}{12}\usefont{T1}{ptm}{m}{n}{\color[rgb]{0,.56,0}shear deformations: $m=0$}%
}}}
\put(4726,-1456){\makebox(0,0)[lb]{\smash{\fontsize{10}{12}\usefont{T1}{ptm}{m}{n}{\color[rgb]{0,.56,0}or e.g.\ {\sl simple shear}}%
}}}
\put(3566,-481){\makebox(0,0)[lb]{\smash{\fontsize{10}{12}\usefont{T1}{ptm}{m}{n}{\color[rgb]{.56,0,.56}$m=\frac13$}%
}}}
\put(3561,-661){\makebox(0,0)[lb]{\smash{\fontsize{10}{12}\usefont{T1}{ptm}{m}{n}{\color[rgb]{.56,0,.56}$\f F={\rm diag}(\lambda, \lambda^{\frac13}, \lambda^{-\frac43})$}%
}}}
\end{picture}%

%% file: InvarEbene_MoRi_Stab_1.pdf_t
\begin{picture}(0,0)%
\includegraphics{InvarEbene_MoRi_Stab_1.pdf}%
\end{picture}%
\setlength{\unitlength}{4144sp}%
\begin{picture}(3956,2915)(860,-3314)
\put(991,-1726){\rotatebox{90.0}{\makebox(0,0)[lb]{\smash{\fontsize{10}{12}\usefont{T1}{ptm}{m}{n}{\color[rgb]{0,0,0}$I_2$}%
}}}}
\put(2971,-3256){\makebox(0,0)[lb]{\smash{\fontsize{10}{12}\usefont{T1}{ptm}{m}{n}{\color[rgb]{0,0,0}$I_1$}%
}}}
\end{picture}%

%% file: InvarEbene_MoRi_Stab_2.pdf_t
\begin{picture}(0,0)%
\includegraphics{InvarEbene_MoRi_Stab_2.pdf}%
\end{picture}%
\setlength{\unitlength}{4144sp}%
\begin{picture}(3956,2915)(860,-3314)
\put(991,-1726){\rotatebox{90.0}{\makebox(0,0)[lb]{\smash{\fontsize{10}{12}\usefont{T1}{ptm}{m}{n}{\color[rgb]{0,0,0}$I_2$}%
}}}}
\put(2971,-3256){\makebox(0,0)[lb]{\smash{\fontsize{10}{12}\usefont{T1}{ptm}{m}{n}{\color[rgb]{0,0,0}$I_1$}%
}}}
\end{picture}%

%% file: InvarEbene_OGDEN_Stab.pdf_t
\begin{picture}(0,0)%
\includegraphics{InvarEbene_OGDEN_Stab.pdf}%
\end{picture}%
\setlength{\unitlength}{4144sp}%
\begin{picture}(3956,2915)(860,-3314)
\put(991,-1726){\rotatebox{90.0}{\makebox(0,0)[lb]{\smash{\fontsize{10}{12}\usefont{T1}{ptm}{m}{n}{\color[rgb]{0,0,0}$I_2$}%
}}}}
\put(2971,-3256){\makebox(0,0)[lb]{\smash{\fontsize{10}{12}\usefont{T1}{ptm}{m}{n}{\color[rgb]{0,0,0}$I_1$}%
}}}
\end{picture}%

%% file: InvarEbene_HENCKYquadr_Stab.pdf_t
\begin{picture}(0,0)%
\includegraphics{InvarEbene_HENCKYquadr_Stab.pdf}%
\end{picture}%
\setlength{\unitlength}{4144sp}%
\begin{picture}(4027,2915)(860,-3314)
\put(991,-1726){\rotatebox{90.0}{\makebox(0,0)[lb]{\smash{\fontsize{10}{12}\usefont{T1}{ptm}{m}{n}{\color[rgb]{0,0,0}$I_2$}%
}}}}
\put(2971,-3256){\makebox(0,0)[lb]{\smash{\fontsize{10}{12}\usefont{T1}{ptm}{m}{n}{\color[rgb]{0,0,0}$I_1$}%
}}}
\end{picture}%

%% file: InvarEbene_Yeoh_Stab__ABAQ.pdf_t
\begin{picture}(0,0)%
\includegraphics{InvarEbene_Yeoh_Stab__ABAQ.pdf}%
\end{picture}%
\setlength{\unitlength}{3522sp}%
\begin{picture}(5699,3070)(860,-3314)
\put(991,-1726){\rotatebox{90.0}{\makebox(0,0)[lb]{\smash{\fontsize{10}{12}\usefont{T1}{ptm}{m}{n}{\color[rgb]{0,0,0}$I_2$}%
}}}}
\put(2971,-3256){\makebox(0,0)[lb]{\smash{\fontsize{10}{12}\usefont{T1}{ptm}{m}{n}{\color[rgb]{0,0,0}$I_1$}%
}}}
\put(2701,-2491){\makebox(0,0)[lb]{\smash{\fontsize{10}{12}\usefont{T1}{ptm}{m}{n}{\color[rgb]{0,0,1}limitation uniaxial deformation: $m=-\frac12$}%
}}}
\put(2701,-2671){\makebox(0,0)[lb]{\smash{\fontsize{9}{10.8}\usefont{T1}{ptm}{m}{n}{\color[rgb]{0,0,1}$\f F={\rm diag}(\lambda, 1/\sqrt{\lambda}, 1/\sqrt{\lambda})$}%
}}}
\put(1486,-376){\makebox(0,0)[lb]{\smash{\fontsize{10}{12}\usefont{T1}{ptm}{m}{n}{\color[rgb]{0,0,1}biaxial deformation}%
}}}
\put(1486,-556){\makebox(0,0)[lb]{\smash{\fontsize{10}{12}\usefont{T1}{ptm}{m}{n}{\color[rgb]{0,0,1}$m=1$}%
}}}
\put(1486,-736){\makebox(0,0)[lb]{\smash{\fontsize{9}{10.8}\usefont{T1}{ptm}{m}{n}{\color[rgb]{0,0,1}$\f F={\rm diag}(\lambda, \lambda, \lambda^{-2})$}%
}}}
\put(4006,-1276){\makebox(0,0)[lb]{\smash{\fontsize{10}{12}\usefont{T1}{ptm}{m}{n}{\color[rgb]{0,.56,0}{\sl pure shear} with $\f F={\rm diag}(\lambda,1,\lambda^{-1})$}%
}}}
\put(4006,-1096){\makebox(0,0)[lb]{\smash{\fontsize{10}{12}\usefont{T1}{ptm}{m}{n}{\color[rgb]{0,.56,0}shear deformations: $m=0$}%
}}}
\put(4726,-1456){\makebox(0,0)[lb]{\smash{\fontsize{10}{12}\usefont{T1}{ptm}{m}{n}{\color[rgb]{0,.56,0}or e.g.\ {\sl simple shear}}%
}}}
\end{picture}%

%% file: InvarEbene_Yeoh_Stab__ABAQ_ganzeEbene.pdf_t
\begin{picture}(0,0)%
\includegraphics{InvarEbene_Yeoh_Stab__ABAQ_ganzeEbene.pdf}%
\end{picture}%
\setlength{\unitlength}{3522sp}%
\begin{picture}(3956,2915)(860,-3314)
\put(991,-1726){\rotatebox{90.0}{\makebox(0,0)[lb]{\smash{\fontsize{10}{12}\usefont{T1}{ptm}{m}{n}{\color[rgb]{0,0,0}$I_2$}%
}}}}
\put(2971,-3256){\makebox(0,0)[lb]{\smash{\fontsize{10}{12}\usefont{T1}{ptm}{m}{n}{\color[rgb]{0,0,0}$I_1$}%
}}}
\end{picture}%

%% file: invar_random.pdf_t
\begin{picture}(0,0)%
\includegraphics{invar_random.pdf}%
\end{picture}%
\setlength{\unitlength}{4144sp}%
\begin{picture}(3851,2949)(1036,-3348)
\put(1253,-1839){\rotatebox{90.0}{\makebox(0,0)[lb]{\smash{\fontsize{10}{12}\usefont{T1}{ptm}{m}{n}{\color[rgb]{0,0,0}$I_2$}%
}}}}
\put(3221,-3290){\makebox(0,0)[lb]{\smash{\fontsize{10}{12}\usefont{T1}{ptm}{m}{n}{\color[rgb]{0,0,0}$I_1$}%
}}}
\end{picture}%

%% file: invar_plot_m.pdf_t
\begin{picture}(0,0)%
\includegraphics{invar_plot_m.pdf}%
\end{picture}%
\setlength{\unitlength}{4144sp}%
\begin{picture}(3851,2949)(1036,-3348)
\put(1253,-1839){\rotatebox{90.0}{\makebox(0,0)[lb]{\smash{\fontsize{10}{12}\usefont{T1}{ptm}{m}{n}{\color[rgb]{0,0,0}$I_2$}%
}}}}
\put(3221,-3290){\makebox(0,0)[lb]{\smash{\fontsize{10}{12}\usefont{T1}{ptm}{m}{n}{\color[rgb]{0,0,0}$I_1$}%
}}}
\put(3061,-2401){\rotatebox{359.9}{\makebox(0,0)[lb]{\smash{\fontsize{10}{12}\usefont{T1}{ptm}{m}{n}{\color[rgb]{0,0,0}uniaxial tension}%
}}}}
\put(2341,-1006){\rotatebox{359.9}{\makebox(0,0)[lb]{\smash{\fontsize{10}{12}\usefont{T1}{ptm}{m}{n}{\color[rgb]{0,0,0}biaxial}%
}}}}
\put(2296,-1186){\rotatebox{359.9}{\makebox(0,0)[lb]{\smash{\fontsize{10}{12}\usefont{T1}{ptm}{m}{n}{\color[rgb]{0,0,0}$m=1$}%
}}}}
\put(3646,-2221){\rotatebox{359.9}{\makebox(0,0)[lb]{\smash{\fontsize{10}{12}\usefont{T1}{ptm}{m}{n}{\color[rgb]{0,0,0}$m=-\frac12$}%
}}}}
\put(2881,-1726){\rotatebox{359.9}{\makebox(0,0)[lb]{\smash{\fontsize{10}{12}\usefont{T1}{ptm}{m}{n}{\color[rgb]{0,0,0}$m=4$}%
}}}}
\put(3466,-1006){\rotatebox{359.9}{\makebox(0,0)[lb]{\smash{\fontsize{10}{12}\usefont{T1}{ptm}{m}{n}{\color[rgb]{0,0,0}$m=5$}%
}}}}
\put(1916,-1546){\rotatebox{359.9}{\makebox(0,0)[lb]{\smash{\fontsize{10}{12}\usefont{T1}{ptm}{m}{n}{\color[rgb]{0,0,0}$m=-3$}%
}}}}
\put(3316,-1316){\rotatebox{359.9}{\makebox(0,0)[lb]{\smash{\fontsize{10}{12}\usefont{T1}{ptm}{m}{n}{\color[rgb]{0,0,0}$m=3$}%
}}}}
\put(4566,-1696){\rotatebox{359.9}{\makebox(0,0)[lb]{\smash{\fontsize{10}{12}\usefont{T1}{ptm}{m}{n}{\color[rgb]{0,0,0}$m=-\frac23$}%
}}}}
\put(3786,-1371){\rotatebox{359.9}{\makebox(0,0)[lb]{\smash{\fontsize{10}{12}\usefont{T1}{ptm}{m}{n}{\color[rgb]{0,0,0}$m=-\frac45$}%
}}}}
\end{picture}%

%% file: InvarEbene_MoRi_ext_HESSE_HILL_c2-0p2.pdf_t
\begin{picture}(0,0)%
\includegraphics{InvarEbene_MoRi_ext_HESSE_HILL_c2-0p2.pdf}%
\end{picture}%
\setlength{\unitlength}{3315sp}%
\begin{picture}(3992,2915)(860,-3314)
\put(991,-1726){\rotatebox{90.0}{\makebox(0,0)[lb]{\smash{\fontsize{10}{12}\usefont{T1}{ptm}{m}{n}{\color[rgb]{0,0,0}$I_2$}%
}}}}
\put(2971,-3256){\makebox(0,0)[lb]{\smash{\fontsize{10}{12}\usefont{T1}{ptm}{m}{n}{\color[rgb]{0,0,0}$I_1$}%
}}}
\end{picture}%

%% file: HESSE_projection.pdf_t
\begin{picture}(0,0)%
\includegraphics{HESSE_projection.pdf}%
\end{picture}%
\setlength{\unitlength}{2072sp}%
\begin{picture}(5219,2013)(1059,-2825)
\put(2971,-1816){\makebox(0,0)[lb]{\smash{\fontsize{12}{14.4}\usefont{T1}{ptm}{m}{n}{\color[rgb]{0,0,0}$\f v$}%
}}}
\put(3826,-1366){\makebox(0,0)[lb]{\smash{\fontsize{12}{14.4}\usefont{T1}{ptm}{m}{n}{\color[rgb]{0,0,0}$\f n$}%
}}}
\put(6031,-2221){\makebox(0,0)[lb]{\smash{\fontsize{12}{14.4}\usefont{T1}{ptm}{m}{n}{\color[rgb]{0,0,0}$\pb\eta$}%
}}}
\put(4861,-2761){\makebox(0,0)[lb]{\smash{\fontsize{12}{14.4}\usefont{T1}{ptm}{m}{n}{\color[rgb]{0,0,0}$\f u$}%
}}}
\end{picture}%

%% file: InvarEbene_MoRi_ext_HESSE_proj_2x2_c2-0p2.pdf_t
\begin{picture}(0,0)%
\includegraphics{InvarEbene_MoRi_ext_HESSE_proj_2x2_c2-0p2.pdf}%
\end{picture}%
\setlength{\unitlength}{3315sp}%
\begin{picture}(3992,2915)(860,-3314)
\put(991,-1726){\rotatebox{90.0}{\makebox(0,0)[lb]{\smash{\fontsize{10}{12}\usefont{T1}{ptm}{m}{n}{\color[rgb]{0,0,0}$I_2$}%
}}}}
\put(2971,-3256){\makebox(0,0)[lb]{\smash{\fontsize{10}{12}\usefont{T1}{ptm}{m}{n}{\color[rgb]{0,0,0}$I_1$}%
}}}
\end{picture}%

%% file: InvarEbene_MoRi_D_ABAQ.pdf_t
\begin{picture}(0,0)%
\includegraphics{InvarEbene_MoRi_D_ABAQ.pdf}%
\end{picture}%
\setlength{\unitlength}{3315sp}%
\begin{picture}(3992,2915)(860,-3314)
\put(991,-1726){\rotatebox{90.0}{\makebox(0,0)[lb]{\smash{\fontsize{10}{12}\usefont{T1}{ptm}{m}{n}{\color[rgb]{0,0,0}$I_2$}%
}}}}
\put(2971,-3256){\makebox(0,0)[lb]{\smash{\fontsize{10}{12}\usefont{T1}{ptm}{m}{n}{\color[rgb]{0,0,0}$I_1$}%
}}}
\end{picture}%

%% file: InvarEbene_MoRi_Stab__summary_c2=-0p4.pdf_t
\begin{picture}(0,0)%
\includegraphics{InvarEbene_MoRi_Stab__summary_c2=-0p4.pdf}%
\end{picture}%
\setlength{\unitlength}{1865sp}%
\begin{picture}(12011,3005)(860,-3404)
\put(991,-1726){\rotatebox{90.0}{\makebox(0,0)[lb]{\smash{\fontsize{10}{12}\usefont{T1}{ptm}{m}{n}{\color[rgb]{0,0,0}$I_2$}%
}}}}
\put(10981,-3346){\makebox(0,0)[lb]{\smash{\fontsize{10}{12}\usefont{T1}{ptm}{m}{n}{\color[rgb]{0,0,0}$I_1$}%
}}}
\put(6931,-3346){\makebox(0,0)[lb]{\smash{\fontsize{10}{12}\usefont{T1}{ptm}{m}{n}{\color[rgb]{0,0,0}$I_1$}%
}}}
\put(2971,-3346){\makebox(0,0)[lb]{\smash{\fontsize{10}{12}\usefont{T1}{ptm}{m}{n}{\color[rgb]{0,0,0}$I_1$}%
}}}
\put(1081,-3346){\makebox(0,0)[lb]{\smash{\fontsize{10}{12}\usefont{T1}{ptm}{m}{n}{\color[rgb]{0,0,0}(a)}%
}}}
\put(5041,-3346){\makebox(0,0)[lb]{\smash{\fontsize{10}{12}\usefont{T1}{ptm}{m}{n}{\color[rgb]{0,0,0}(b)}%
}}}
\put(9136,-3346){\makebox(0,0)[lb]{\smash{\fontsize{10}{12}\usefont{T1}{ptm}{m}{n}{\color[rgb]{0,0,0}(c)}%
}}}
\end{picture}%

%% file: InvarEbene_Yeoh_Stab__summary.pdf_t
\begin{picture}(0,0)%
\includegraphics{InvarEbene_Yeoh_Stab__summary.pdf}%
\end{picture}%
\setlength{\unitlength}{1865sp}%
\begin{picture}(12011,3005)(860,-3404)
\put(991,-1726){\rotatebox{90.0}{\makebox(0,0)[lb]{\smash{\fontsize{10}{12}\usefont{T1}{ptm}{m}{n}{\color[rgb]{0,0,0}$I_2$}%
}}}}
\put(10981,-3346){\makebox(0,0)[lb]{\smash{\fontsize{10}{12}\usefont{T1}{ptm}{m}{n}{\color[rgb]{0,0,0}$I_1$}%
}}}
\put(6931,-3346){\makebox(0,0)[lb]{\smash{\fontsize{10}{12}\usefont{T1}{ptm}{m}{n}{\color[rgb]{0,0,0}$I_1$}%
}}}
\put(2971,-3346){\makebox(0,0)[lb]{\smash{\fontsize{10}{12}\usefont{T1}{ptm}{m}{n}{\color[rgb]{0,0,0}$I_1$}%
}}}
\put(1081,-3346){\makebox(0,0)[lb]{\smash{\fontsize{10}{12}\usefont{T1}{ptm}{m}{n}{\color[rgb]{0,0,0}(a)}%
}}}
\put(5041,-3346){\makebox(0,0)[lb]{\smash{\fontsize{10}{12}\usefont{T1}{ptm}{m}{n}{\color[rgb]{0,0,0}(b)}%
}}}
\put(9136,-3346){\makebox(0,0)[lb]{\smash{\fontsize{10}{12}\usefont{T1}{ptm}{m}{n}{\color[rgb]{0,0,0}(c)}%
}}}
\end{picture}%

%% file: SiSh_logb12.pdf_t
\begin{picture}(0,0)%
\includegraphics{SiSh_logb12.pdf}%
\end{picture}%
\setlength{\unitlength}{3315sp}%
\begin{picture}(4767,2919)(803,-3318)
\put(2971,-3256){\makebox(0,0)[lb]{\smash{\fontsize{10}{12}\usefont{T1}{ptm}{m}{n}{\color[rgb]{0,0,0}shear $\gamma=\tan\alpha$}%
}}}
\put(946,-2356){\rotatebox{90.0}{\makebox(0,0)[lb]{\smash{\fontsize{10}{12}\usefont{T1}{ptm}{m}{n}{\color[rgb]{0,0,0}shear comp.\ $(\log\f b)_{12}$}%
}}}}
\put(4141,-2311){\makebox(0,0)[lb]{\smash{\fontsize{10}{12}\usefont{T1}{ptm}{m}{n}{\color[rgb]{0,0,0}$x_1$}%
}}}
\put(3691,-2536){\makebox(0,0)[lb]{\smash{\fontsize{10}{12}\usefont{T1}{ptm}{m}{n}{\color[rgb]{0,0,0}$1$}%
}}}
\put(2791,-1321){\makebox(0,0)[lb]{\smash{\fontsize{10}{12}\usefont{T1}{ptm}{m}{n}{\color[rgb]{0,0,0}$x_2$}%
}}}
\put(3916,-1501){\makebox(0,0)[lb]{\smash{\fontsize{10}{12}\usefont{T1}{ptm}{m}{n}{\color[rgb]{0,0,0}$\gamma$}%
}}}
\put(2836,-1681){\makebox(0,0)[lb]{\smash{\fontsize{10}{12}\usefont{T1}{ptm}{m}{n}{\color[rgb]{0,0,0}$1$}%
}}}
\put(3106,-1951){\makebox(0,0)[lb]{\smash{\fontsize{9}{10.8}\usefont{T1}{ptm}{m}{n}{\color[rgb]{0,0,0}$\alpha$}%
}}}
\end{picture}%